\documentclass{amsart}
\usepackage{psfrag}
\psfrag{A}{\fontsize{24}{24}$\psi_0^\ve$}
  \psfrag{B}{\fontsize{24}{24}$\ov{\psi_0^\ve}$}

\psfrag{1}{\fontsize{16}{16}$k_n$}
\psfrag{2}{\fontsize{16}{16}$k_{n-1}$}
\psfrag{3}{\fontsize{16}{16}$k_{n-2}$}
\psfrag{n}{\fontsize{16}{16}$k_1$}
\psfrag{o}{\fontsize{24}{24}$\mathcal{O}_\varepsilon$}
\psfrag{j}{\fontsize{20}{20}$t$}
\psfrag{q}{\fontsize{16}{16}$v + \xi / 2$}
\psfrag{w}{\fontsize{16}{16}$v - \xi / 2$}
\psfrag{r}{\fontsize{16}{16}$v - \xi / 2 + \sum_{l=1}^n k_l$}
\psfrag{e}{\fontsize{16}{16}$v + \xi / 2 + \sum_{l=1}^n k_l$}
\psfrag{c}{\fontsize{16}{16}$v + \xi / 2$}
\psfrag{d}{\fontsize{16}{16}$v - \xi / 2$}

\newtheorem{theorem}{Theorem}[section]
\newtheorem{lemma}[theorem]{Lemma}

\theoremstyle{definition}

\newtheorem{cor}[theorem]{Corollary}

\theoremstyle{remark}

\newtheorem{remark}[theorem]{Remark}
\newcommand{\bes}{{\begin{split}}}
\newcommand{\ees}{{\end{split}}}
\newcommand{\bees}{{\begin{equation}\begin{split}}}
\newcommand{\es}{{\end{split}\end{equation}}}
\newcommand{\erre}{{\mathbb R}}

\newcommand{\natu}{{\mathbb N}}

\newcommand{\comple}{{\mathbb C}}

\newcommand{\bea}{\begin{eqnarray}}

\newcommand{\eea}{\end{eqnarray}}

\newcommand{\be}{\begin{equation}}

\newcommand{\ee}{\end{equation}}

\newcommand{\n}{\noindent}

\newcommand{\f}{\frac}

\newcommand{\al}{\alpha}

\newcommand{\ve}{\varepsilon}

\newcommand{\om}{\omega}

\newcommand{\ov}{\overline}

\newcommand{\nnn} {\nonumber}

\newcommand{\tal}{\widetilde \alpha}

\newcommand{\mc}{\mathcal}

\numberwithin{equation}{section}

\begin{document}
\large

\title{Rate of decoherence for an electron \\
weakly coupled to a phonon gas}


\author{Riccardo Adami}

\address{R. Adami: Dipartimento di Matematica e Applicazioni,
Universit\`a di Milano Bicocca.}
\curraddr{Via R. Cozzi 53, 20125 Milano, Italy.}
\email{riccardo.adami@unimib.it}



\thanks{R.A. is grateful to the Mathematisches Institut der
Ludwig-Maximilians-Universit\"at
   M\"unchen for the kind hospitality, and to 
the Istituto Nazionale di Alta Matematica for providing financial
support through the ``Borsa per l'Estero INdAM'' fellowship}


\author{L\'aszl\'o Erd\H{o}s}

\address{L. Erd\H{o}s:
Mathematisches Institut der Ludwig-Maximilians-Universit\"at
   M\"unchen.}
\curraddr{39 Theresienstr., D-80333, M\"unchen, Germany.}
\email{lerdos@mathematik.uni-muenchen.de}

\date{February 8, 2008}


\keywords{}

\begin{abstract} We study the dynamics of an electron weakly coupled
   to a phonon gas.
The initial state of the electron is  the superposition of two spatially
localized  distant
bumps moving towards each other, and the phonons are in a
thermal state. We investigate the
dynamics of the system in the kinetic
regime and show that
the time evolution makes
the non-diagonal terms of the density matrix
of the electron
decay, destroying the interference between the two bumps.
We show that such a damping effect is exponential in time, and the
related decay rate is proportional to the total
scattering cross section of the  electron-phonon interaction.
\end{abstract}

\maketitle

\section{Introduction and result}\label{sec:intro}
  Quantum interference of a state can be destroyed via
  interactions with an external environment. This phenomenon is
called decoherence and it is
recognized to be
responsible for the transition from quantum superpositions to
classical probability measures. This feature makes it relevant in the 
field
of quantum information, where  the superposition of states
allows for parallel computation and therefore it must be preserved.

Rigorous results on  decoherence have so far concerned
systems with a discrete energy spectrum  or decoherence induced
by practically instantaneous scattering at time zero.
For the first model, we mention that the decoherence of
an electron trapped in a harmonic oscillator
and interacting with an electromagnetic field in the
dipole approximation was studied in \cite{ds}. For the second situation, 
in
\cite{dft, afft1, afft2}
it was
showed that decoherence is the most relevant effect in the dynamics
of many heavy particles in an environment of light
particles, when the mass ratio is small.
For an extensive physical
introduction to decoherence we refer to \cite{gk,omnes}.

Here we investigate the occurrence of decoherence in the
dynamics of an electron weakly coupled
to a phonon field. The energy spectrum of the electron is
continuous and the phonons constantly bombard the electron
along its time evolution. We will study the problem in
the kinetic regime with a weak coupling.
We introduce a scale parameter $\ve$ and analyze the time evolution at
large space and time scales of order $\ve^{-1}$.
If $(x,t)$ denote the original (microscopic) space and time coordinates, 
then
we introduce the macroscopic space and time coordinates $(X, T)$ by
\begin{equation*}
\begin{array} {cc}  T := \ve t, & X : = \ve x . \end{array}
\end{equation*}
The coupling constant between the electron and phonon field will be given 
by
$\lambda = \sqrt{\ve}$.

On the macroscopic space and time scales,
the dynamics of the electron in such a limit is described
by a linear Boltzmann equation, as proven in \cite{e}.
  In this paper we prove that, in addition
to such  behavior,  on the microscopic scale the
quantum interference remains detectable and
decays exponentially in time with a rate
growing with the strength of the electron-phonon interaction.


To be more specific, we choose a special initial state
that exhibits a strong interference pattern without phonon interaction.
We assume that initially
the electron lies in the superposition of two distant
macroscopic wave packets (``bumps'') in $\erre^d$ ($d \geq 3$)
  moving
towards each other, namely
\begin{equation} \begin{split}
\psi_0^\ve (x) & := \ \psi_{0,+}^\ve (x)
+ \psi_{0,-}^\ve (x)  \\
\psi_{0,+}^\ve (x) & := \ \ve^{d/2} f (\ve x + Q) e^{i P \cdot x} \\
  \psi_{0,-}^\ve (x) & := \ \ve^{d/2} f (\ve x - Q) e^{-i P \cdot x}
\label{initalelec0}
\end{split}
\end{equation}
where $f$ is  $L^2$-normalized and satisfies
the following hypothesis of regularity
\be \label{regf}
f \in H^{\f {3d} 2 + 6} (\erre^d).
\ee
The vectors $P, Q\in \erre^d$ are  parallel, i.e.
$P \cdot Q = |P| |Q|$ and assume $P \neq 0 $.
The wave packets $\psi_{0,-}^\ve$ and $\psi_{0,+}^\ve$
are localized around the macroscopic points $Q$ and $-Q$
and they move towards each other by a momentum $P$.
The function $f$ describes the macroscopic envelope
of the bumps. We remark that $\psi_0^\ve$ is not normalized in $L^2$,
but it is easily seen that $\lim_{\ve \to 0} \| \psi_0^\ve \| = \sqrt 2$,
so we will always consider $\ve$ sufficiently small to guarantee
$ 1 \ \leq \ \| \psi_0^\ve \|_2   \ \leq \ 2. $

To explain our result, we preliminarily examine what happens if
the interaction between electron and field is absent.
In this case the state of the electron $\psi_t^\ve = \psi_{t, 
\rm{free}}^\ve$
evolves according to the free Schr\"odinger equation
$i\partial_t\psi_t
= -\frac{1}{2}\Delta\psi_t$.
The two wave packets maximally overlap at time
$\bar t \ = \ \ve^{-1} |Q|/ |P|$, when the probability density of finding
the electron at $x$, apart from the normalization factor, is given by
$$ | \psi_{\bar t, \rm{free}}^{\ve} (x)
  |^2 \ = \ 2  ( 1 + \cos (2 P \cdot x) )
\left|  \ve^{d/2}
\left( e^{i \ve^2 \bar t \f \Delta 2} f \right) (\ve x)
\right|^2. $$

The shape of the bumps has undergone a semiclassical
change and their supports are shifted
to the origin. If the state were classical, the probability densities
of the bumps should add up, yielding the factor of 2.
The
cosine term carries the typical interference fringes and therefore the
information that the electron lies in a quantum superposition of two
wave packets.

Let us remark the presence of two spatial scales in
$ | \psi_{\bar t, \rm{free}}^{\ve} (x)
  |^2$:
the envelope of the wave function is spread  on a length of order
$\ve^{-1}$, while the interference pattern manifests itself on a
length scale of order $1$.
The size of the interference fringes can be quantified
evaluating the Fourier transform of $ | \psi_{\bar t, \rm{free}}^{\ve}|^2$
  and then
performing the limit $\ve \rightarrow 0$. One easily sees
\begin{equation}\label{ptilde}
\lim_{\ve \rightarrow 0}
\int dx \, e^{2 i \widetilde P \cdot x}  | \psi_{\bar t, 
\rm{free}}^{\ve}(x)
  |^2 \ =
\left\{
\begin{array}{lll} 2 & \hbox{if $\widetilde P=0$} \\
1 & \hbox{if $\widetilde P= \pm P$}\\
0 & \hbox{otherwise.}
\end{array}
\right.
  \end{equation}

If, instead of being isolated,
  the electron interacts from the beginning with
a phonon field that initially lies in a thermal state, then
the interference
fringes are expected
to be damped by a factor exponentially decreasing in the
collision time in macroscopic units,  i.e. with $|Q|/|P|$.

Indeed, we prove that
$$
\lim_{\ve \rightarrow 0}
\int dx \, e^{\pm 2 i P \cdot x}   \rho_{\bar t}^{\ve} (x)
   \ = \ e^{- \f { \sigma_P |Q| }{|P|}}
$$
where $\sigma_P$, given in \eqref{sigmap},
  is the total scattering cross section for an electron
with velocity $P$ interacting with the phonon field, and
$ \rho_{\bar t}^{\ve} (x)$ is the probability density
of finding the interacting electron at $x$ at time $\bar t$.
The interacting Hamiltonian will be defined in Section \ref{sec:model}.
  Our method easily extends to more general initial states
whose free evolution exhibits interference fringes.
  For simplicity we discuss only the particular case given
by \eqref{initalelec0}.

To state our result in a more precise way, we use the Wigner
function formalism instead of the wave-function description
  of the state of the electron. At time zero,
the Wigner function of the electron reads
\begin{equation} \label{initalwig0}
W_0 (x,v) \ : = \  \int_{\erre^d} \f {dy} {(2 \pi)^d} \, e^{-i v
   \cdot y} \psi_0^\ve ( x + y / 2) {\ov{\psi_0^\ve ( x - y / 2)}}\; .
\end{equation}
According to the decomposition of $\psi_0^\ve$ given in
(\ref{initalelec0}),
formula (\ref{initalwig0}) yields a decomposition in four terms
for the
Wigner function of the initial state
\begin{eqnarray*}
W_0 \ =  \sum_{\alpha, \alpha^\prime \in \{ \pm \}} W_{0,{\alpha
\alpha^\prime}}, & &
  W_{0,{\alpha
\alpha^\prime}} (x,v) \ : = \  \int_{\erre^d} \f {dy} {(2 \pi)^d}
\, e^{-i v \cdot y} \psi_{0, \alpha}^\ve ( x + y / 2)
{\ov{\psi_{0, \alpha^\prime}^\ve ( x - y / 2)}} \; ,
\end{eqnarray*}
where the so-called diagonal terms $W_{0,++}$ and $W_{0,--}$
  represent
an electron in the state $ \psi_{0, +}^\ve$ and $ \psi_{0, -}^\ve$
respectively, and the non-diagonal terms
$W_{0,+-}$ and $W_{0,-+}$ represent the interference between
the states $ \psi_{0, +}^\ve$ and $ \psi_{0, -}^\ve$.

Since the time evolution of the Wigner function is linear, one can study
separately the evolution of diagonal terms
$ W_{++}(t)$ and $ W_{--}(t)$ and of non-diagonal ones
$ W_{+-}(t)$ and $ W_{-+}(t)$.
We will use the notations $W_{\pm,\pm, \rm{free}}(t)$
and $W_{\pm,\pm}(t)$ for the corresponding components of the
Wigner transform of the state at time $t$ under the free and the 
interacting
evolutions, respectively. The interacting evolution will be defined in
Section \ref{sec:model}.
We remark that after tracing out
the phonon degrees of freedom, the electron will be
in a mixed state and thus $W_{\pm,\pm}(t)$
will be Wigner functions of density matrices.

In order to easily define the phonon field
operators in a rigorous way we consider the system  confined in a
$d$-dimensional box $\Lambda_L = \left[ - L/2,  L/2 \right]^d$ and then
perform the thermodynamic limit $L \rightarrow \infty$. Such a limit
is taken after the phonon trace but
before the scaling limit $\ve \rightarrow 0$, and the result
is uniform in $L$.
The parameter $L$ will be suppressed in the notation of $W_{\pm,\pm}(t)$.
Before performing the thermodynamic limit
$L \rightarrow \infty$, any integral
in the space variables will be thought of as one over
  the volume $\Lambda$, and any
integral in the momentum variable is actually a summation over
$\Lambda^\star = (L^{-1}\mathbb Z)^d$.
For the computation of the
thermodynamic limit we refer to section 4.5 in \cite{e}.
The free evolution will always be considered in the infinite volume
to avoid taking the unnecessary thermodynamic limit in
  $W_{\pm,\pm, \rm{free}}(t)$.

We will test the evolved Wigner functions against observables,
  $J(x,v)= J_\ve(x,v)$ that may scale with $\ve$ to detect
the interference fringes.
Let $\langle J, W\rangle =\int \ov{J(x,v)}W(x,v)d x d v$
denote the expectation
value of the observable $J$
on a state given by the Wigner function $W$.

We will always assume that
\begin{equation}
    \| J \| : = \sup_\ve \int \sup_v |\widehat J_\ve(\xi, v)| d\xi < 
\infty\; ,
\label{obscond}
\end{equation}
where we use the convention that the hat on functions defined
on the electronic phase space denotes the Fourier transform in
the space variable only, i.e.
$$
    \widehat J_\ve(\xi, v)= \frac{1}{(2\pi)^{d/2}} \int e^{-i\xi\cdot x} 
J_\ve(x,v) dx \; .
$$

In the following Theorem  \ref{teorema} we compare the
  kinetic limit for
$ W_{+-}(t)$ and $ W_{-+}(t)$ with the corresponding terms
under the free evolution.
With these  notations, we have the following main theorem:
\begin{theorem} \label{teorema}  Let the initial state of the electron
be given by (\ref{initalelec0}) with $P\neq 0$ and fix a macroscopic time 
$T$.
At any time $t$, let
    $ W_{+-,\rm{free}} (t)$  be
a non-diagonal component of the Wigner function
  of a non-interacting (``isolated'') electron evolved
under the free evolution.
Let   $ W_{+-} (t)$ be
  the analogous non-diagonal component
of the Wigner function describing the electron at time $t$
interacting with a phonon field through the interaction Hamiltonian
(\ref{interaction}).
Then, for any observable $J_\ve$ satisfying \eqref{obscond}, we have
\begin{eqnarray}
\lim_{\ve \rightarrow 0}\left| \lim_{L\to \infty}
\langle J_\ve, W_{+-} (\ve^{-1} T)
\rangle ~ - ~ e^{- {T} \sigma_P }
\langle J_\ve, W_{+-,\rm{free}} (\ve^{-1} T)
\rangle\right|   & = &0 \label{result}
\end{eqnarray}
where
$\sigma_P$, given in \eqref{sigmap},
  is the total cross section for an electron with momentum $P$
in the phonon field.
\end{theorem}

  To detect the destruction  of the
interference fringes, the two-scale structure
of $ W_{+-}(t)$ and $ W_{-+}(t)$ obliges us
to test  $ W_{+-}(t)$ and $ W_{-+}(t)$ against
observables endowed with the same two-scale structure.
A possible class of such observables $J_\ve$
will be given in Section \ref{sec:obs}.
In particular,  for these observables
one easily obtains that $\lim_{\ve \to 0} \langle J_\ve, W_{+-,\rm{free}} 
(\ve^{-1} T)
\rangle$ exists.  The precise statement is formulated
in Corollary \ref{cor:cor}.

  The evolution of the diagonal terms, $W_{++}$ and  $W_{--}$,
on the macroscopic scales
is given by the linear Boltzmann equation,
\begin{equation}
    (\partial_T +\nabla e(V)\cdot \nabla_X) F_T(X,V) =
\int \sigma(V,U) F_T(X,U) dU -
  \Big(\int \sigma(U,V) d U\Big)F_T(X,V)
\label{linbol}
\end{equation}
with collision kernel $\sigma(U,V)$
and free dispersion relation $e(V) = \frac{1}{2}V^2$.
The two terms on the right hand side
are called the gain and loss terms, respectively.
For the precise  statement, see \cite{e}.

Theorem \ref{teorema} describes the evolution of the off-diagonal terms.
The physical scenario can be explained as follows: in the kinetic
limit, the two initial bumps evolve independently as classical
phase space probability densities
obeying the linear Boltzmann equation (\ref{linbol}).
In addition to that, there is an interference
term that equals the corresponding term for the free case, except for a 
damping
factor, exponential in time. In other words, the interference term
evolves according to a Boltzmann equation without gain term.

From the technical point of view,  we
expand the time evolution into a Duhamel sum, trace out the degrees
of  freedom of the phonon and lastly perform the scaling limit.
This is conveniently made using Feynman graph expansions,
similarly to \cite{ey} and \cite{e}.
The novelty in our model is that the relevant graphs are no longer given 
by the
ladder terms: these ones vanish in the limit and
  the resulting dynamics is given by the renormalized free evolution.

The paper is organized as follows.
In Section \ref{sec:not} we fix some basic notation and conventions; in 
Section \ref{sec:ass} we
state regularity assumptions on the dispersion relations and on the form
factor of the interaction, and give a lemma on some estimates that will
be used repeatedly along the paper. In Section \ref{sec:model}
  we define the model and  the
scaling, and specify the assumptions on the initial state and the 
observables.
In Section \ref{sec:known} we explain the link with
the result provided in \cite{e}.
In Section \ref{sec:ladder} we prove that the contribution of the
ladder graphs is
negligible in the kinetic limit, and in Section \ref{sec:main}
we compute the contribution
due to renormalized free propagators.
The paper ends by an appendix containing the
proof of the lemma stated in Section \ref{sec:ass}.

\section{Notation and conventions}\label{sec:not}

We model the electron as a spinless particle, so the state space for
the electron is ${\mc H}_{\rm e} : = L^2_{\rm{per}} (\Lambda_L)$.
The electron dispersion
relation is denoted by $e(k)$, $k \in \erre^d$, and the Hamiltonian
$H_{\rm e}$ of a free electron acts on ${\mc H}_{\rm e}$ as follows
\begin{equation}
\label{freeel}
H_{\rm e} \ = \ e (-i \nabla).
\end{equation}
One can think of the classical dispersion relation
$e(p) = \frac {|p|^2} {2M}$
or the pseudo-relativistic one $e(p) = \sqrt{|p|^2+
   M^2}$, where $M$ is the mass of the electron.

Let $\gamma$ be the density operator representing the state
of the electron;
$\gamma$ is a positive operator on $L^2 (\erre^d)$
and  its  operator
kernel is denoted by $\gamma (x,y)$.
Let $\widehat \gamma$ be the Fourier transform of
$\gamma$ as an operator, i.e. its kernel is given by
\begin{equation*}
\widehat \gamma(p,u) \ : = \ \int \f{dx dy}{(2 \pi)^{d}}
e^{-ip\cdot x+iu \cdot y}
\gamma (x,y).
\end{equation*}

The Wigner transform
of $\gamma$ is defined as
\begin{equation}
\label{wignertrans}
W_\gamma (x,v) \ = \ \int e^{-i v \cdot y} 
\gamma \left( x + \f y 2, x - \f
   y 2 \right) \, \f {dy} {(2 \pi)^d}.
\end{equation}
$W_\gamma$ is a real function defined on the classical phase space of the
electron. We recall the convention that the hat on functions defined
on the phase space of the electron denotes Fourier transform in the
space
variable only, i.e.
\begin{equation}
\label{spacevar}
\widehat W_\gamma (\xi,v) \ = \ \int  \f{dx}{(2 \pi)^{\f d 2}}
e^{-i \xi\cdot x} \, W_\gamma (x,v)
\ = \ ( 2 \pi)^{-\f d 2}
\widehat \gamma \left(v + \f \xi 2, v - \f \xi 2 \right).
\end{equation}
The rescaled (macroscopic) Wigner function is defined as follows
\begin{equation*}
W^\ve_\gamma (X,V) \ = \ \ve^{-d} W_\gamma \left( \f X \ve, V \right).
\end{equation*}
If the density matrix $\gamma$ is the orthogonal projection on the
space spanned by the funtion $\varphi$ (i.e. the electron lies in a pure
state), then we denote the associated Wigner function by $W_\varphi$. We 
will
sometimes use symbols like $W(t)$, without explicitly referring
to a density matrix $\gamma$. In such cases, we will always mean the
Wigner function of the electron at time $t$, i.e.
the Wigner transform of the reduced density matrix for the electron
$\gamma_{{\rm{e}},t}$ to be defined in \eqref{reduced}.

\medskip

The pure
states of  the phonon field are represented by vectors in the
bosonic Fock space
${\mc {H}}_{\rm{ph}} = \oplus_{n=0}^\infty
\left[L^2(\Lambda) \right]^{\otimes_s n}$, where $\otimes_s n$ is the
$n$-fold symmetrized tensor product.
We introduce
the phonon creation and
annihilation operators, $a_k^\dagger$, $a_k$, with momentum $k$,
satisfying the usual commutation relations
$$
     [ a_k, a_{k'}^\dagger]=\delta(k-k')\; .
$$
The number operator of the phonons in mode $k$ is
$$ {\mc N_k}  \ : = \ a^\dagger_k a_k.
$$

The Hamiltonian of a free phonon field reads
\begin{equation}
\label{freeph}
H_{\rm{ph}} \ = \ \int dk \, \omega (k) \, {\mc N}_k.
\end{equation}
Here $\omega (k)$ is the dispersion relation for phonons.
The state of thermal
equilibrium  for a phonon
field at inverse temperature $\beta$ and chemical potential
$\mu$ is given by the density operator
\begin{equation}
\label{gibbs}
\gamma_{\rm{ph}} \ : = Z^{-1}\Big[ {e^{- \beta H_{\rm{ph}}+ \mu \int N_k
       dk}} \Big], \qquad \mbox{with} \quad Z \
: = \ {\mbox{Tr}}_{\mc{H}_{\rm{ph}}} \left(    {e^{-
\beta H_{\rm{ph}} + \mu \int  N_k
       dk}}\right) .
\end{equation}
The expected number of phonons in the mode $k$ reads
\begin{equation*}
{\mc N} (k) : = \mbox{Tr}_{{\mc H}_{\rm{ph}}} ( \gamma_{\rm{ph}} N_k ) \
= \ \f {e^{-\beta \omega (k) + \mu}} {1 -e^{-\beta \omega (k) + \mu}}.
\end{equation*}

The symbol $C$ will
denote various positive constants arising in  estimates.

\section{Assumptions}
\label{sec:ass}
In the following we will
  use extensively the notation $\langle x \rangle : = (x^2 + 1)^{\f
   12}$ and the estimates
\begin{equation} \label{lrangle}
\begin{split}
\langle x + y \rangle \ \leq & \, C \langle x \rangle \langle  y \rangle \\
\langle x + y \rangle^{-1} \ \leq & \,
C \langle x \rangle^{-1} \langle  y \rangle.
\end{split}
\end{equation}

  The dispersion relations $e(\cdot)$, $\omega (\cdot)$ for
electrons and phonons are assumed to be:
\begin{itemize}
\item spherically symmetric;
\item decaying for large $k$ up to the $2d^{\rm{th}}$ derivative:
\begin{equation} \label{decdisprel}
\begin{split}
| \nabla^l e (k) | \ \leq \ C ( 1 + \langle k \rangle^{2-l} ), & \, \
\, l =
0,1, 2, \dots, 2d ; \\
| \nabla^l \omega (k) | \ \leq \ C ( 1 + \langle k \rangle^{2-l} ),
  & \, \ \, l =
0,1, 2, \dots, 2d .
\end{split}
\end{equation}
\end{itemize}
We will often consider the functions
\begin{equation} \label{Phi}
\Phi_\pm (p,k) : = e (k + p ) \pm \omega (k).
\end{equation}
We assume that the
following relations hold
\begin{eqnarray}
\lim_{k \rightarrow \infty} \Phi_\pm (p,k) & = & \infty \label{largek}\\
\exists \ C_1, C_2 \in \erre \ & {\mbox{ s.t. }} & 0 < C_1 \leq
{\mbox{Hess}}_k   \Phi_\pm (p,k) \leq C_2 . \label{hessian}
\end{eqnarray}
Let us consider the ``thick level sets''
$$ E_\pm (p, \theta, \delta ) \ : = \ \{ k: \ |  \Phi_\pm (p,k) -
\theta | \leq \delta \}$$
for the functions $\Phi_\pm$. From (\ref{decdisprel}), (\ref{largek}),
(\ref{hessian}) one can conclude that there exist $\tilde C$, $\tilde
\rho$   such that for any $\delta, \rho \leq \tilde \rho$
\begin{equation} \label{intersection}
\sup_{p,q,\theta} | E_\pm (p, \theta, \delta) \cap B (q, \rho) | \leq
\tilde C \delta \rho^{d-1},
\end{equation}
where $|\cdot|$ denotes the Lebesgue measure, and $ B (q, \rho)$ is
the ball of radius $\rho$ centered in $q \in \erre^d$. We require
a further condition on the intersection of two such sets: there exist
$\tilde \rho, C_3 > 0$ such that for any $\delta_1, \delta_2, \rho
\leq \tilde \rho$, $p_1, p_2 \in \erre^d$, $\theta_1, \theta_2 \in
\erre$
\begin{equation}
\sup_q | E_\pm (p_1, \theta_1, \delta_1) \cap
E_\pm (p_2, \theta_2, \delta_2)
  \cap B (q, \rho) | \leq \f{C_3 \delta_1 \delta_2 \rho^{d-2}}{|p_1 -
    p_2|}.
\label{transversality}
\end{equation}
The listed conditions are fulfilled for a classical or a
pseudo-relativistic electron
if $\|  \nabla^2 \omega \|_\infty$ is sufficiently small.

To ensure that the density operator for the phonon field is trace
class we assume
\begin{equation*}
\inf_k \omega (k) - \mu \beta^{-1} \geq C > 0 .
\end{equation*}
The electron and the phonon field will be coupled via an
interaction  form factor $F (k):\erre^d\to \erre$.
It is
assumed to be real and symmetric,
namely
$$ F (k) = F (-k) = \overline{F(k)} $$
and it has a fast decay up to $2d$-derivatives
\begin{equation}
\max_{l=0, \dots, 2d} \left| \nabla_k^l F (k) \right| \ \leq \ C
\langle k \rangle ^{-2d-12}.
\label{qdecay}
\end{equation}
For $\sigma = \pm 1$ we define the functions
\begin{equation*}
L (k, \sigma) : = \left| {F} (k) \right|^2 \left(  {\mc N} (k)
+ \f {\sigma + 1} 2 \right)
\end{equation*}
and
\begin{equation}
\label{calel}
{\mc L} (k) : = L (k, 1) =  \left| {F} (k) \right|^2 \left(  {\mc N} (k)
+ 1 \right).
\end{equation}
Obviously, $L (k, \pm 1) \leq {\mc L} (k)$, and
$\mc L$ shares with $F$ the decay estimate (\ref{qdecay}).

These conditions on $e(k)$, $\omega(k)$ and $F(k)$ ensure
the following key estimates that we will subsequently use:
\begin{lemma} \label{threees}
For the functions $\mc L$ defined in (\ref{calel}), and $\Phi_\pm$ defined 
in
(\ref{Phi}), the following estimates hold
\begin{eqnarray}
\sup_{p, \theta} \int_{\erre^d} \frac {{\mc L}(k) \, dk}
{\left|  \theta - \Phi_\pm (p,k) + i \eta
\right|^{m+1}} & \leq & C \eta^{-m}, \quad m > 0;  \label{uno} \\
\int_{\erre^d} \frac {{\mc L}(k) \, dk } {\left| \theta - \Phi_\pm (p,k) + 
i \eta
\right| | \widetilde \theta - \Phi_\pm (u,k) - i \eta
| \langle p \rangle  \langle u \rangle}
  & \leq & \frac {C (\log^\star \eta)^2} { | p - u |_\star  \langle \theta
    \rangle^
\f 1 2  \langle \widetilde \theta \rangle^\f 1 2}; \label{due}\\
\sup_v \int_{\erre} \frac {d\alpha}
{\left| \alpha - e (v_2 + \xi / 2)  + i \eta
\right|  \langle \alpha \rangle}
  & \leq & {C \log^\star \eta}; \label{tre}
\end{eqnarray}
where $\eta > 0$ and we used the notation
$f^\star = \max (1,|f|)$, $f_\star = \min (1,|f| + \eta)$.
\end{lemma}
\n
The new estimate compared with \cite{e} is \eqref{due}, whose
  proof is given in the appendix.

In the proof of Theorem \ref{teorema} we will use the following
inequality too:
\begin{equation}
\label{e-4-7}
\sup_v \left| \sum_{\sigma \in \{ \pm \}} \int e^{-i s \Phi_\sigma
    (p,k)} L (k, \sigma) \, dk
\right| \ \leq \ \f C {\langle s \rangle^{\f d 2}}
\end{equation}
The proof of \eqref{e-4-7} is in Lemma 4.1 of \cite{e}.

\section{The model}\label{sec:model}
\subsection{Hamiltonian}
The
dynamics of the system consisting in an electron interacting with a
phonon field is generated by the { Hamiltonian}
\begin{equation*}
H \ = \ H_{\rm e} \otimes \mathbb{I}_{{\mathcal{H}}_{\rm {ph}}} +
\mathbb{I}_{{\mathcal{H}}_{\rm e}}
  \otimes H_{\rm{ph}} + H_{I}.
\end{equation*}
In any tensor product the first factor
acts on
${\mathcal{H}}_{\rm e}$ and the second on ${\mathcal{H}}_{\rm {ph}}$.
The free Hamiltonians $H_{\rm e}$ and $ H_{\rm{ph}}$ have been defined
in (\ref{freeel}) and (\ref{freeph}).
The interaction Hamiltonian $H_I$ is defined by
\begin{equation}
H_I \ = \ i \lambda \int dk \, F(k) \, ( e^{-ik \cdot x } a^\dagger_k  -
  e^{ik \cdot x } a_k). \label{interaction}
\end{equation}
As stated in Section \ref{sec:intro} we will always assume $\lambda = 
\sqrt \ve$.

We assume that initially the electron and
the phonons are independent and their
initial state is
  represented by the density operator
\begin{equation*}
\Gamma_0 \ : = \ \gamma_{{\rm{e}},0} \otimes  \gamma_{\rm{ph}}
\end{equation*}
where $\gamma_{{\rm{e}},0}$ is an electronic density matrix and
$\gamma_{\rm{ph}}$ is the phonon thermal state defined in (\ref{gibbs}).
The time evolution of $\Gamma_t$ is described by
\begin{equation*}
i \partial_t \Gamma_t \ = \ [ H, \Gamma_t ]
\end{equation*}
with $\Gamma_0$ as initial data, or, equivalently
\[
\Gamma_t \ = \ e^{-iHt} \Gamma_0  e^{iHt} .
\]

\subsection{Initial state of the electron}
The initial state of the electron is represented by the wave function
\eqref{initalelec0}.
The density operator corresponding
to the wave function
$\psi_0^\ve$ is the one-dimensional projection
\begin{equation*}
\gamma_{{\rm{e}},0}\ = \ |\psi_0^\ve \rangle\langle \psi_0^\ve|
\end{equation*}
with integral kernel
\begin{equation*}
\gamma_{{\rm{e}},0}(x,y)\ = \ \psi_0^\ve (x) {\ov{\psi_0^\ve (y)}}.
\end{equation*}
According to (\ref{initalelec0}), the initial density operator
for the electron can be split in four terms
\begin{equation}
\gamma_{{\rm{e}},0}  \ = \gamma_{{\rm{e}},0,++} + \gamma_{{\rm{e}},0,--} +
\gamma_{{\rm{e}},0,+-} + \gamma_{{\rm{e}},0,-+}
\label{gammadec}
\end{equation}
where
\begin{equation}
\gamma_{{\rm{e}},0,\alpha \alpha^\prime} \ = \  |\psi_{0, \alpha}^\ve
\rangle\langle\psi_{0, \alpha^\prime}^\ve|
\label{decgamma}
\end{equation}
with $\alpha, \alpha^\prime \in \{ + , - \}$. We stress once again
that, whereas
$\gamma_{{\rm{e}},0,++}$ and $\gamma_{{\rm{e}},0,--}$
describe an
electron in the state $\psi_{0,+}^{\ve}$ and  $\psi_{0,-}^{\ve}$,
respectively,
the terms $\gamma_{{\rm{e}},0,+-}$ and $\gamma_{{\rm{e}},0,-+}$ do not
represent any physical state: in particular, as operators they are not
positive.  They represent the interference between the
states  $\psi_{0,+}^{\ve}$ and  $\psi_{0,-}^{\ve}$.

According to (\ref{gammadec}), the Wigner  transform
of the initial state for the electron
can be written as
\begin{equation} \label{wigdec}
      W_{\psi_0^\ve} = W_{0,++}  + W_{0,--} + W_{0,+-} + W_{0,-+}
\end{equation}
where
\begin{equation*} \begin{split}
W_{0,++} (x,v) & = \ {W}_f
\left(\ve x + Q, \f {v -P} \ve \right) \\
W_{0,--} (x,v) & = \ {W}_f \left(\ve x - Q, \f {v + P} \ve \right)
  \\
W_{0,+-} (x,v) & = \  e^{2 i P \cdot x}  e^{\f {2 i v \cdot Q} \ve}
{W}_f \left(\ve x , \f v \ve \right)
  \\
W_{0,-+} (x,v) & = \ e^{-2 i P \cdot x}  e^{-\f {2 i v \cdot Q} \ve}
{W}_f \left(\ve x , \f v \ve \right)
\end{split} \end{equation*}
as one can immediately verify by \eqref{wignertrans}.

Recalling the convention adopted in (\ref{spacevar})
\begin{equation} \begin{split} \label{initnond}
\widehat W_{0,+-} (\xi,v) \ = & \left( \sqrt{2 \pi}\ve\right)^{-d}
e^{2i \f {v\cdot Q} \ve} \widehat f \left( \f {v} \ve+ \f {\xi-2P} {2 \ve} 
\right)
  {\ov {\widehat f \left( \f {v} \ve -  \f {\xi-2P} {2 \ve} \right)}}
\\
\widehat W_{0,-+} (\xi,v) \ = & \left( \sqrt{2 \pi}\ve\right)^{-d}
e^{-2i \f {v \cdot Q} \ve} \widehat f
\left( \f {v} \ve + \f {\xi+2P} {2 \ve} \right)
  {\ov {\widehat f \left(\f {v} \ve    - \f {\xi+2P} {2 \ve}\right)}}.
\end{split} \end{equation}
All components of $W_0$ depend on $\ve$,
  but this is
omitted in notation.

\subsection{Time evolution of the electron}
The state  of the electron at time $t$ can be described by the marginal 
(or
reduced density operator)
\begin{equation}
\label{reduced}
\gamma_{{\rm{e}},t} : = {\rm{Tr}_{\mc{H}_{\rm{ph}}}} \Gamma_t
\end{equation}
or, equivalently, by the Wigner
function
$$ W (t;x,v) \ = \ \int \f{dy}{(2 \pi)^d} e^{-i v \cdot y }
   \gamma_{{\rm{e}},t} \left(x+\frac y 2, x - \frac y 2\right).
$$
Since the time evolution is linear, the initial decompositions
(\ref{gammadec})  and (\ref{wigdec}) propagate
at any time $t$:
\begin{equation} \label{t-dec} \begin{split}
\gamma_{{\rm{e}},t}  \ = \ & \gamma_{{\rm{e}},t, ++} + 
\gamma_{{\rm{e}},t,--}
  + \gamma_{{\rm{e}},t,+-}  + \gamma_{{\rm{e}},t,-+} \\
W (t) \ = \ & W_{++}(t) +   W_{--} (t)    +   W_{+-} (t)    +   W_{-+} (t)
\end{split} \end{equation}
Again, the first two terms in the decomposition are called diagonal, and
the other two terms non-diagonal.

\subsection{Observables} \label{sec:obs}
We will use the notation $\mc{O}$ for the observables
considered as operators and their Wigner transforms
will be denoted by $J= W_{\mc{O}}$. The
  expectation value of an observable is given by
$$
{\rm{Tr}}\; \mc{O}^\star  \gamma_{\rm{e}}
=   \int \ov{J(x,v)} W(x,v)dx dv  = \int \ov{ \widehat J_\ve (u,v)}
\widehat W \left( u,v \right)
\, du dv  .
$$
We need observables  capable to resolve the
size of the interference fringes. Therefore we
consider observables $\mc{O}={\mc O}_\ve$
endowed with a two-scale structure in the space
variables but we always assume the condition \eqref{obscond} on $J_\ve$.
This implies that
\be
\label{stimoss}| \! | \!  | \mc{O} | \! | \!  |^2 : =  \sup_\ve
\| {\mathcal O}_\ve^\star {\mathcal O}_\ve \|
< \infty
\ee
since the operator norm of $\mc{O}^\star \mc{O}$ is
bounded by $\| J\|^2$.

In order to select observables that actually detect the exponential
decay of the interference fringes,
we provide sufficient conditions on $J_\ve$
for $\lim_{\ve \rightarrow 0} \langle J_\ve, W_{+-, \rm{free}}
(\ve^{-1} T ) \rangle$ to exist.
We assume that the Wigner transform of $\mc{O}_\ve$
has the structure
\begin{equation}\label{obsfourier}
J_\ve (x,v) \ = \ A(\ve x, v) b (x)
\end{equation}
where $A$
describes the macroscopic profile of the observable and $b$ encodes
the short-scale structure.

First we consider for simplicity
  the case of Newtonian dispersion relation,
i.e. $e(v) = \f 1 2 v^2$. One has
$$ W_{+-, \rm{free}} (\ve^{-1} T ;x,v) \ = \ e^{2i P \cdot x} e^{2 i \f v 
\ve
\cdot ( Q - P T)} W_f \left( \ve x - vT,  \f v \ve  \right)
$$
Due to the regularity hypothesis \eqref{regf}, $W_f $
belongs to
$L^1 (\erre^{2d})$, and then
$\| W_{+-, \rm{free}} (\ve^{-1} T ) \|_1 = \|  W_f  \|_1$.
Besides, due to \eqref{obscond},
it is clear that $J_\ve \in  L^\infty (\erre^{2d})$ with a uniform bound
in $\ve$.

After an elementary change of variables, one gets
$$   \langle J_\ve, W_{+-, \rm{free}}
(\ve^{-1} T ) \rangle \ = \ \int dx \, dv \, \ov{A} (x + \ve vT, \ve v)
\ov{b} \left( \f x  \ve +vT   \right) e^{2 i P \cdot \left( \f x \ve
+ vT \right)}
e^{2 i v \cdot ( Q - PT) } W_f (x,v)$$
and the limit for $\ve \rightarrow 0$ can be investigated using the 
dominated
convergence theorem, assuming that $A$ and $b$ are
bounded and continuous functions. One finds that:
  \begin{itemize}
\item if
$ \lim_{|x| \to \infty} b (x)  = 0$,
then $\lim_{\ve \to 0}  \langle J_\ve, W_{+-, \rm{free}}
(\ve^{-1} T ) \rangle$ vanishes. In this case the observable
does not resolve the interference fringes, even though the limit of
the expectation value exists;
\item if
$ \lim_{|x| \to \infty} b (x) e^{-2 i P \cdot x} =: c_b \neq 0$,
then
\be \label{limexpec}
  \lim_{\ve \to 0}  \langle J_\ve, W_{+-, \rm{free}}
(\ve^{-1} T ) \rangle = \ov{c_b} \int dx \, dv \, \ov{A (x,0)}
  e^{2 i v \cdot ( Q - PT)} W_f (x,v)
\ee
\end{itemize}

In general, the short-scale factor $b$ must exhibit
the periodicity of the interference
fringes. We require 
the distributional Fourier transform
$\widehat b(\xi)$ to be a complex measure
with a finite total variation. If it has a non-trivial Dirac delta
component with frequency $2P$, i.e.
\be
\lim_{\tau \to 0+0} \frac{1}{(2\pi)^{d/2}}\int_{| \xi - 2P| \leq \tau}
\widehat b (\xi)
\, d \xi = : c_b \neq 0,
\label{bdelta}
\ee
then the interference fringes in $W_{+-}$ can
be detected by $J_\ve$. For the other off-diagonal term, $W_{-+}$,
one needs a non-trivial component with frequency $-2P$.
Working in Fourier space requires more conditions on $A$,
for simplicity we assume that $A$ is a Schwartz function.

In the case of  a more general dispersion relation $e(v)$
the Wigner transform of the free evolution, $W_{+-, {\rm{free}}} (t)$,
is given by
\begin{equation} \label{freeWt}
{\widehat W}_{+-, {\rm{free}}} (t; \xi, v) = {(\sqrt{2 \pi} \ve)}^{-d}
  e^{-it [e
     (v + \xi / 2) - e ( v - \xi / 2) ]} e^{2 i v \cdot \f Q \ve} \widehat 
f \left(
     \f v \ve + \f {\xi -2P} {2\ve} \right) \ov{ \widehat f \left(
      \f v \ve - \f {\xi -2P} {2\ve} \right)}
\end{equation}
in Fourier space.
After a
change of variable one gets
\begin{equation*} \begin{split}
\langle J_\ve, W_{+-, \rm{free}}
(\ve^{-1} T ) \rangle \ = & (2 \pi)^{-d} \int d\xi \, dv \, d\zeta \,
\ov{\widehat A} \left( \xi + \f {2 P - \zeta} \ve, \ve v \right)
\ov{ \widehat b}
(\zeta) e^{-iT\ve^{-1}[ e (\ve v + \ve \xi/2 + P ) -  e (\ve v - \ve \xi/2 
- P
   )]} \\ & \times
e^{ 2i v \cdot Q} \widehat f \left( v + \f \xi 2 \right)
\ov{\widehat f \left( v -  \f \xi 2 \right)}.
\end{split} \end{equation*}
Again, by using dominated convergence, we have
$$ \lim_{\ve \to 0}
\langle J_\ve, W_{+-, \rm{free}}
(\ve^{-1} T ) \rangle \ = \ (2 \pi)^{-d/2}
\ov{c_b} \int d \xi \, dv \,
\ov{\widehat A ( \xi,
0)} e^{2 i v \cdot ( Q - T \nabla e (P))}
    \widehat f \left( v + \f \xi 2 \right)
\ov{\widehat f \left( v -  \f \xi 2 \right)} $$
that, recalling \eqref{spacevar}, generalizes
  \eqref{limexpec} to a generic dispersion relation.

In summarizing, we have proved the following
\begin{cor} \label{cor:cor}
  {\em  Assume the  hypotheses of Theorem \ref{teorema} and let  $J_\ve 
(x,v) = A (\ve x, v) b (x)$. Let either
\begin{itemize}
\item[i)] $e(v)$ be quadratic, $A\in C_b(\erre^{2d})$, $b\in C_b(\erre^d)$ 
and
$ c_b:=\lim_{|x| \to \infty} b (x) e^{-2 i P \cdot x}$; or
\item[ii)] $A\in {\mathcal S}(\erre^{2d})$ and $\widehat b$
be a measure with finite total variation with $c_b$ being
the coefficient of the Dirac delta component at $2P$
of this measure (see
  \eqref{bdelta}).
\end{itemize}
Then,
  \begin{eqnarray*}
\lim_{\ve \rightarrow 0}  \lim_{L\to\infty}
\langle J_\ve, W_{+-} (\ve^{-1} T)
\rangle & = & e^{- {T} \sigma_P } \lim_{\ve \rightarrow 0}
\langle J_\ve, W_{+-,\rm{free}} (\ve^{-1} T)
\rangle \\  & = &  e^{- {T} \sigma_P }
\ov{c_b} \int dx \, dv \, \ov{A (x,0)}
  e^{2 i v \cdot ( Q - T \nabla e (P))} W_f (x,v) \; .
\end{eqnarray*}}
\end{cor}

\n
The hypotheses
cover the case
$ J_\ve (x,v)=  e^{2i\widetilde P \cdot x} $ for the quadratic dispersion 
relation
(see \eqref{ptilde}).

\section{Known results}\label{sec:known}
The proof partially follows the arguments in  \cite{e}.
To avoid duplicating them but still keep the present paper
relatively self-contained, here we summarize some
notations and results from \cite{e}.

\subsection{The main term of $\gamma_{\rm{e},t}$}\label{sec:ladder1}
We consider the density matrix of the whole system
\[
\Gamma_t \ = \ e^{-iHt} \Gamma_0  e^{iHt}
\]
and expand the propagator $e^{-iHt}$ into a
Duhamel sum up to the order
\[
N_0 \ = \ \f{2.2 \log \ve} {\log | \log \ve |}.
\]
After performing the phonon trace and defining the reduced density matrix
of the electron as
$$ \gamma_{{\rm{e}},t} \ = \ {\rm{Tr}_{\rm{ph}}} \left(
e^{-iHt} \Gamma_0 e^{iHt} \right),$$
$ \gamma_{{\rm{e}},t}$ is
decomposed  into the sum of a main term and an error term
\be \label{main-err}
\gamma_{{\rm{e}},t} \ = \ \gamma_{K}^{\rm{main}} (t) +
\gamma_{K}^{\rm{err}} (t)
\ee
where $K\in\natu$ is an arbitrary integer.
For the precise definitions of $ \gamma_{K}^{\rm{main}} (t)$ and
$\gamma_{K}^{\rm{err}} (t)$ see formula (2.18) in \cite{e},
but their explicit form will not be needed
here. We need only to know that
\be \label{lim}
\limsup_{K \rightarrow \infty}
\limsup_{\ve \rightarrow 0} \limsup_{L \rightarrow \infty}
\left| {\rm{Tr}} \; \gamma_K^{\rm{err}} (\ve^{-1}T) {\mc{O}}_\ve^\star
\right| \ = \ 0,
\ee

\n
i.e. that the estimate (2.24) of \cite{e} holds in our case as well.
To arrive at this result, the argument in \cite{e} used only
assumption (1.26) in \cite{e}  on the initial state and
assumption (2.21) in \cite{e} on the observable. In our case, the first
assumption is guaranteed by the hypothesis \eqref{regf};
  the second one is   inequality
(\ref{stimoss}). For the same reason, Lemma 3.1, Proposition
10.1 and Lemma 10.2  from \cite{e} also hold
in our case.

To evaluate $ \gamma_K^{\rm{main}} (t)$ we use Proposition
10.1 and Lemma 10.2 in \cite{e}. Results therein can be resumed
as follows
\be \label{limsup} \begin{split} &
\limsup_{\ve \to 0}
 \Bigg|  \lim_{L \rightarrow \infty} \langle J_\ve ,
   W_{\gamma_K^{\rm{main}}(\ve^{-1}T)}  \rangle   \\ &
- \sum_{N, \widetilde N = 0}^{K-1} \sum_{n=0}^{\min (N, \widetilde N)}
\sum_{\footnotesize{\begin{array}{c}
\underline m = (m_0, \dots, m_n) \in \natu^{n+1} \\
N = n +
2 \sum_{j=0}^n m_j \end{array}}}
\sum_{\footnotesize{\begin{array}{c}
\underline {\widetilde {m}} = (\widetilde m_0, \dots, \widetilde m_n)
\in \natu^{n+1} \\
\widetilde N = n +
2 \sum_{j=0}^n \widetilde m_j \end{array}}}
C^*_{\underline m, \widetilde{\underline m}, {\rm{id}}}(\ve^{-1}T) \Bigg| = 0
\end{split} \ee
where, for any
$\underline m, \underline {\widetilde m} \in \natu^{n+1}$, $t > 0$
  we defined
\bea
C^*_{\underline m, \widetilde{\underline m}, {\rm{id}}}(t)
& : = &
\left( {2 \pi}\right)^{-2}  \lambda^{2n + 2 |
\underline m |+ 2 |\underline {\widetilde m}|}
\sum_{\stackrel{\sigma_j \in
   \{ \pm \}} {j = 1 , \dots , n}}
   \int d\xi  \, dv \,
\left(\prod_{j=1}^n dk_j \right) \, \ov{\widehat J_\ve \left( \xi,
   v + \sum_{l=1}^n
   k_l \right)} \nonumber \\ & & \times
\widehat W_{\psi_0^\ve} (\xi,v)
  \left(
   \prod_{j=1}^n L (k_j, \sigma_j) \right)
e^{2 t \eta} \int_\erre d\alpha \, {e^{-it\alpha}}  \left(
   \prod_{j=0}^n R_j^{m_j + 1}  {\Upsilon}_j^{m_j} \right)
\nnn \\ & &  \times \int_\erre d\tal \, {e^{it\tal}}  \left(
   \prod_{j=0}^n {\widetilde R}_j^{{\widetilde m}_j + 1}
{\widetilde {\Upsilon
     }}_j^{\widetilde m_j} \right)  \nnn \\ & &
\label{ladderterm}
\end{eqnarray}
Here $\eta > 0$,
$\underline k = (k_1, \dots, k_n) \in \erre^{nd}$, $|
\underline m| = \sum_{j=0}^n m_j $,  $|
\underline {\widetilde{m}} | = \sum_{j=0}^n \widetilde{m}_j $
and to define $R_j$ and $\Upsilon_j$
we first introduce
\begin{equation} \label{r-upsilon} \begin{split}
   R(\alpha, v, z) & = \ \frac{1}{\alpha - e(v) - z} \\
\Upsilon_\eta ( \alpha, v) & = \ \sum_{\sigma = \pm} \int \f{ L (k,
   \sigma) dk} {\alpha - e (v + k) - \sigma \om (k) + i \eta}
\end{split} \end{equation}
for $\alpha\in \erre$, $v\in \erre^d$, $z\in \comple$, and $\eta > 0$,
and then we set
\begin{eqnarray*}
    R_j &:= & R\left(\alpha, v + \sum_{l=j+1}^n
k_l + \f \xi 2, \sum_{l= j+1}^n \sigma_l
\omega (k_l)-i \eta \right) \\
   \widetilde R_j & := & R\left(\widetilde \alpha, v + \sum_{l=j+1}^n
k_l - \f \xi 2, \sum_{l= j+1}^n \sigma_l
\omega (k_l)+i \eta \right) \\
{\Upsilon}_j  & : = & {\Upsilon}_\eta
\left( \alpha -  \sum_{l= j+1}^n  \sigma_l
\omega (k_l) , v + \sum_{l=j+1}^n
k_l + \f \xi 2 \right) \\
{\widetilde {\Upsilon}_j }  & : = &
{\overline{\Upsilon}_\eta}  \left( \widetilde
\alpha -  \sum_{l= j+1}^n  \sigma_l
\omega (k_l), v + \sum_{l=j+1}^n
k_l - \f \xi 2 \right)
\end{eqnarray*}
Formula \eqref{ladderterm} holds for any $\eta>0$;
later we will choose $\eta= \ve$.

The following estimates will be used in the next section
\begin{equation} \begin{split} \label{useful}
\sup_{p,\alpha,\eta} \left| {\Upsilon}_\eta
\left(\alpha , p \right) \right| \leq C, ~  \, & \qquad ~
| R_j |^{m_j + 1} \leq \eta^{-m_j} | R_j |, \\
\sup_{p, \alpha, \eta} \left( |\nabla_p \Upsilon_\eta ( \alpha, p )| +
  |\partial_\alpha \Upsilon_\eta ( \alpha, p )| \right. & \left. \! +  \
|\partial_\eta \Upsilon_\eta ( \alpha, p )|
\right) \ \leq \ C \eta^{-1/2}
\end{split} \end{equation}
The first and the third
inequalities
in \eqref{useful} are
  proven in [E] (Lemma 4.1), the second is trivial.

\subsection{Ladder Wigner functions and related Feynman graphs}
We introduce the family of ``ladder Wigner functions''
$W^{\rm{ladder}}_{n, \underline m, \underline
  {\widetilde m}} (t)$, defined implicitly by the relation
\be \label{defladderW}
C^*_{\underline m, \underline {\widetilde m}, \rm{id}} (t) \ = \
\langle J_\ve, W^{\rm{ladder}}_{n, \underline m, \underline
  {\widetilde m}} (t) \rangle.
\ee
The explicit expression of $W^{\rm{ladder}}_{n, \underline m, \underline
  {\widetilde m}} (t)$
in the Fourier space can be easily obtained comparing
\eqref{defladderW} with \eqref{ladderterm}.
The content of (\ref{limsup}) can be expressed saying that,
in the thermodynamic limit,
  the expectation value of
$\mc{O}_\ve$ on $\gamma_K^{\rm{main}} (t)$
can be decomposed in the sum of the expectation values of
$J_\ve$ on the states represented by
$W^{\rm{ladder}}_{n, \underline m, \underline  {\widetilde m}} (t)$.

The function $\widehat W^{\rm{ladder}}_{n, \underline m, \underline
  {\widetilde m}} (t; \xi, v)$ is linear in $\widehat W_{\psi_0^\ve}$,
  thus  it contains two evolved copies of the
electron wavefunction:
  one copy initially lies in the state represented by
$\psi_0^\ve$,
the other in $\ov{\psi_0^\ve}$. In the following we will call them,
for short, the first and the second copy.
In the  time interval $(0,t)$  each copy of the
electron emits and absorbs phonons. An important feature of
the evolution described by $\widehat W^{\rm{ladder}}_{n, \underline m, 
\underline
  {\widetilde m}} (t; \xi, v)$ is that any emitted phonon must be
reabsorbed and vice versa, but a phonon emitted by one copy of the
electron can be
reabsorbed by the other.

This fact permits us to distinguish two different processes: the
exchange of phonons between the two copies of the electron, and the
recollision of the same copy with the same phonon.
In the evolution
described by $\widehat W^{\rm{ladder}}_{n, \underline m, \underline
  {\widetilde m}} (t; \xi, v)$ all possible recollisions are immediate, 
i.e., there
are no further events in the time interval between the emission of a
phonon by a copy of the electron and its reabsorption by the same
copy.
Moreover, exchanged phonons appear in the same order in the time
evolution of the two copies of the wavefunction of the electron.

In (\ref{ladderterm}) there are $n$ exchanged phonons and their
  momenta are labeled $k_1, \dots, k_n$.
The electron propagator between consecutive phonon collisions
is represented
by the factors $R_j$ (for the first copy)
and $\widetilde R_j$ (for the second). Between the
$j$th and the $(j+1)$st exchanges of
  phonons, the first copy of the electron undergoes
  $m_j$ immediate recollisions,
each of them embodied in a factor $ {\Upsilon}_j$
  in (\ref{ladderterm}).
Analogously, the second copy of the electron suffers
  $\widetilde m_j$ immediate recollisions, each of them represented in
(\ref{ladderterm})  by a factor  $ \widetilde {\Upsilon}_j$.

The expectation value of an  observable $\mc O_\ve$ on the
state represented by $\widehat W^{\rm{ladder}}_{n, \underline m, 
\underline
  {\widetilde m}} (t; \xi, v)$
can be represented as a Feynman graph (see Fig. 1).


\begin{figure}
\begin{center}
{}{}\scalebox{0.60}{\includegraphics{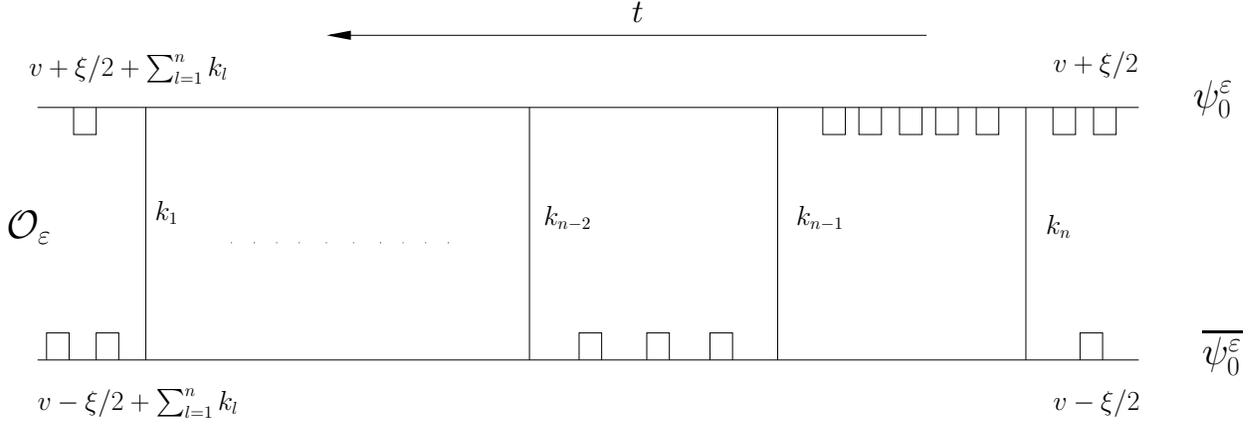}}
\caption{A ladder Feynman graph. The two horizontal
lines represent the time evolution of the two
copies of the electron, and the vertical lines joining the
   two horizontal ones represent the
   exchanges of phonons between them. They are
   labeled by the momenta of the exchanged phonons.
By convention time goes
   from the right to the left and the labeling
of the exchanged phonons
  follows the opposite ordering.
The square loops on the higher horizontal line symbolize the
recollisions in the evolution of the first copy of the electron.
Such loops
have no momentum labeling because in the related factors ${\Upsilon}_j$
  an internal loop momentum integral has been performed.
An analogous interpretation holds
for the square loops in the lower horizontal line.
Here
$m_0 = 1, m_{n-2} = 0, m_{n-1} = 5, m_n =2$ and
$\widetilde m_0 = 2, \widetilde m_{n-2} = 3, \widetilde m_{n-1} = 0,
\widetilde m_n =1$.}
\end{center}
\end{figure}


\subsection{Non-diagonal terms} \label{sec:non-diag}

Following (\ref{t-dec}), the expectation of any
ladder Wigner function can be
decomposed in diagonal and non diagonal terms
$$
    C^*_{\underline m, \widetilde{\underline m}, 
{\rm{id}}}(t)=\sum_{\alpha,\alpha'\in\{ +, -\}}
  C^*_{\alpha\alpha' ,\underline m, \widetilde{\underline m}, {\rm{id}}}(t)
$$
where $C^*_{\alpha\alpha' ,\underline m, \widetilde{\underline m}, 
{\rm{id}}}(t)$
is defined from \eqref{ladderterm} after replacing
$\widehat W_{\psi_0^\ve}$ by $\widehat W_{0,\alpha \alpha^\prime}$.
The corresponding Wigner transforms,
  $W^{\rm{ladder}}_{\alpha \alpha^\prime,n, \underline m, \underline
  {\widetilde m}} (t)$,   are defined implicitly via
\be \label{Wladder}
    C^*_{+-,\underline m, \widetilde{\underline m}, {\rm{id}}}(t)
   = \langle J_\ve, W^{\rm{ladder}}_{\alpha \alpha^\prime,n, \underline m, 
\underline
  {\widetilde m}} (t)\rangle
\ee
and of course,
$$
W^{\rm{ladder}}_{n, \underline m, \underline  {\widetilde m}} (t)
\ = \ \sum_{\alpha,\alpha'\in \{+,-\}}
W^{\rm{ladder}}_{\alpha\alpha',n, \underline m, \underline  {\widetilde 
m}}
(t)
$$

Since both the evolution of density matrices
and
the definition of $\gamma_K^{\rm{main}} (t)$ and $\gamma_K^{\rm{err}} (t)$
are linear in the initial data (see (2.18) and (2.19) in \cite{e}),
  all results listed in Section \ref{sec:ladder1} hold
separately for diagonal and non-diagonal terms, according to
decompositions
\eqref{t-dec}. In particular, formulas \eqref{main-err}, \eqref{lim},
\eqref{limsup} remain valid after replacing $\gamma_{{\rm{e}},t}$,
$\gamma_K^{\rm{main}} (t)$ and $\gamma_K^{\rm{err}} (t)$ with the
corresponding
non-diagonal component $\gamma_{{\rm{e}},t,+-}$,
$\gamma_{K,+-}^{\rm{main}} (t)$ and $\gamma_{K,+-}^{\rm{err}} (t)$,
respectively.

Then, formulas \eqref{lim}, \eqref{limsup}, applied to non-diagonal
terms give
\begin{equation*} 
\begin{split} &
\limsup_{K \to \infty} \limsup_{\ve \rightarrow 0} 
  \Bigg|  \lim_{L \rightarrow \infty}\langle J_\ve ,
   W_{+-}(\ve^{-1}T) \rangle  \\ &
- \sum_{N, \widetilde N = 0}^{K-1} \sum_{n=0}^{\min (N, \widetilde N)}
\sum_{\footnotesize{\begin{array}{c}
\underline m = (m_0, \dots, m_n) \in \natu^{n+1} \\
N = n +
2 \sum_{j=0}^n m_j \end{array}}}
\sum_{\footnotesize{\begin{array}{c}
\underline {\widetilde {m}} = (\widetilde m_0, \dots, \widetilde m_n)
\in \natu^{n+1} \\
\widetilde N = n +
2 \sum_{j=0}^n \widetilde m_j \end{array}}}
C^*_{+-,\underline m, \widetilde{\underline m}, {\rm{id}}}(\ve^{-1}T) \Bigg| = 0 .
\\ \end{split}\end{equation*}

In \cite{e} it is proved that the limit
\[
\lim_{K \rightarrow \infty}
\lim_{\ve \rightarrow 0} \lim_{L \rightarrow \infty}
\left[ {\rm{Tr}} \;  \gamma_{K}^{\rm{main}} (\ve^{-1}T)
{\mc{O}_\ve^\star}
\right]
\]
equals
the expectation value of ${\mc O}_\ve$ on a distribution on the phase 
space
that solves a linear Boltzmann equation, provided that  ${\mc O}_\ve$ has
no short-scale structure (i.e. $b=1$ in \eqref{obsfourier} ).

In the rest of the paper
  we study the same limit of the non-diagonal part
for a two-scale observable  \eqref{obsfourier}.
In the following section we prove that ladder terms with at least one
rung do not
contribute. In contrast, diagrams
with no rungs and pure recollisions give a non vanishing contribution,
characterized by the exponential damping described in Section 
\ref{sec:intro}.
This is the content of Section \ref{sec:main}.

\section{Non-diagonal $n$-rung ladder term with immediate
   recollisions: estimate}\label{sec:ladder}

We estimate the contribution of the generic non-diagonal ladder term
$W_{+-,n,\underline{m},,\underline{\widetilde m} }^{\rm{ladder}}
(t)$ to the expectation value of a two-scale observable $J_\ve$.
\begin{lemma} \label{ladd+-}
Given $n \geq 1$, $\underline{m}, \underline{\widetilde m} \in
\natu^{n+1}$, $P \neq 0$,
let $\widehat W^{\rm{ladder}}_{+-,n,\underline{m},
   \underline{\widetilde m}}$ be defined as in \eqref{Wladder}, and
$J_\ve$ be a two-scale observable satisfying estimate \eqref{obscond}.
Then for any fixed $T>0$
\begin{eqnarray}
\left|\langle J_\ve,
W_{+-,n,\underline{m},,\underline{\widetilde m} }^{\rm{ladder}}
(\ve^{-1} T)
\rangle \right|
& \leq & C ^{n + |
\underline m |+  |\underline {\widetilde m}|} \| J\|
\f {\langle P \rangle^2} { |P|_\star}
(\log^\star \ve)^{4} \ve
  \| f \|^2_{H^2 (\erre^d)},
  \label{letzte1}
\end{eqnarray}
where $\| J\|$ is defined in \eqref{obscond},
$| P |_\star = \min ( 1, | P | + \ve )$, $\log^\star \ve = \max( 1, |\log 
\ve|)$
and the constant $C$ depends on $T$.
\end{lemma}

\begin{proof}
From \eqref{initnond},
  \eqref{ladderterm} and \eqref{Wladder} we have
\begin{eqnarray*}
& &
\langle J_\ve, W_{+-,n,{\underline m},{\underline {\widetilde
     {m}}}}^{\rm{ladder}}
(t) \rangle
\nonumber \\
& =  & \left( {2 \pi}\right)^{-2-\f d 2} \ve^{-d} \lambda^{2n + 2 |
\underline m |+ 2 |\underline {\widetilde m}|} \sum_{\stackrel{\sigma_j 
\in
   \{ \pm \}} {j = 1 , \dots , n}}
   \int d\xi  \, dv \,
\left(\prod_{j=1}^n dk_j \right) \, \ov{\widehat J_\ve \left( \xi,
   v + \sum_{l=1}^n
   k_l \right)} e^{2i \f {v \cdot Q} \ve}
\nonumber
\\ & & \times
\widehat f \left(  \f {v} \ve + \f {\xi-2P}  {2 \ve}
   \right)
\ov{\widehat f \left( \f {v} \ve - \f {\xi-2P} {2 \ve}  \right)} \left(
   \prod_{j=1}^n L (k_j, \sigma_j) \right)
e^{2 t \eta} \int_\erre d\alpha \, {e^{-it\alpha}}  \left(
   \prod_{j=1}^n R_j^{m_j + 1}  {\Upsilon}_j^{m_j} \right)
\nnn \\ & & \times \int_\erre d\tal \, {e^{it\tal}}  \left(
   \prod_{j=1}^n {\widetilde R}_j^{{\widetilde m}_j + 1}  {\widetilde 
{\Upsilon
     }}_j^{\widetilde m_j} \right) \; . \nnn \\ & &
\end{eqnarray*}
To estimate $| \langle J_\ve,
W_{+-,n,{\underline m},{\underline {\widetilde
     {m}}}}^{\rm{ladder}}
(\ve^{-1} T) \rangle |$ we use the first two
inequalities in \eqref{useful}, with $\eta = \ve$,
 and definition \eqref{calel}.
Then
\begin{eqnarray}
& &
\left| \langle
   J_\ve, W_{+-,n,{\underline m},{\underline {\widetilde
     {m}}}}^{\rm{ladder}}
(\ve^{-1} T) \rangle \right|
\nonumber \\
& \leq & \ve^{-d} \lambda^{2n}  ( C \lambda^2 \ve^{-1})^{|
\underline m |+  |\underline {\widetilde m}|}
   \int d\xi  \, dv  \, \left\| \widehat J_\ve \left( \xi,
\cdot \right) \right\|_\infty
\left| \widehat f \left( \f {v} \ve + \f {\xi-2P}  {2 \ve}
    \right) \right| \left|
\widehat f \left( \f {v} \ve - \f {\xi-2P} {2 \ve}  \right)  \right|
\nnn \\ & & \times \sum_{\stackrel{\sigma_j \in
   \{ \pm \}} {j = 1 , \dots , n}} \left(
   \prod_{j=1}^n \int \mc{L} (k_j) \, dk_j \right)
\int_\erre d\alpha   \left(
   \prod_{j=1}^n  | R_j|  \right)
   \int_\erre d\tal \left(
   \prod_{j=1}^n |{\widetilde R}_j| \right) \; .  \nnn \\ & & 
\label{reladder}
\end{eqnarray}
Let us focus on the integral in
$dk_1 \dots dk_n d\al d\tal$.
\begin{eqnarray}
& &   \left( \prod_{j=1}^n \int {\mathcal L} (k_j) \, dk_j \right)
\int_\erre d\alpha   \left(
   \prod_{j=1}^n  | R_j|  \right)
   \int_\erre d\tal \left(
   \prod_{j=1}^n |{\widetilde R}_j| \right) \nnn \\ & = &
\int {d \al d\tal}  | R_n|   |{\widetilde R}_n| \int  dk_n  \,
  {\mathcal L} (k_n) | R_{n-1}|
|{\widetilde R}_{n-1}| \dots  \int  dk_1  \, {\mathcal L} (k_1) | R_{0}|
|{\widetilde R}_{0}| \label{purenlad}
\end{eqnarray}
Indeed, the variable $k_l$ appears in $R_j$ and ${\widetilde R}_{j}$ only
if $j \leq l-1$.  Therefore, we can integrate in (\ref{purenlad})
following the order $k_1, \dots, k_n$.
The integrals in $k_j$ with $1 \leq j \leq n-1$ are estimated
by standard Cauchy-Schwarz inequality
\begin{eqnarray}
\int { {\mathcal L} (k_j) |R_{j-1}| |\widetilde R_{j-1}| \,  dk_j}
\nnn \  \leq \ C \int { {\mathcal L} (k_j) |R_{j-1}|^2 \,  dk_j}
\, + \, C  \int { {\mathcal L} (k_j) |\widetilde R_{j-1}|^2 \,  dk_j}
\nnn & \leq &  C \ve^{-1} \nnn \\ \label{standardcsi}
\end{eqnarray}
where we exploited estimate \eqref{uno}, with
$\eta = \ve$  and $m\ge 1$. We finally obtain the factor
$C^{n-1} \ve^{-n+1}$.

The integral in $k_n$ is estimated using (\ref{due}) with $\theta =
\alpha$, $\widetilde \theta = \widetilde \alpha$, $p = v + \xi / 2$,
$\widetilde p = v - \xi / 2$, $\eta=\ve$. We obtain
\begin{eqnarray}
\int \frac{ {\mathcal L} (k_n) \,  dk_n} {| \alpha - e (v +   \xi/2
+  k_n ) - \sigma_n \omega(k_n) + i \ve |
| \tal - e (v - \xi/2 + k_n ) -  \sigma_n
\omega(k_n) - i \ve |}
& & \nnn \\ \leq \ C \frac{(\log^\star \ve)^2 \langle v + \xi /2 \rangle
\langle v - \xi /2 \rangle}{ | \xi |_\star \langle \al
   \rangle^{1/2}
  \langle \tal \rangle^{1/2}} & &
\nnn \\ \label{kn}
\end{eqnarray}
where, according to the notation used in Lemma \ref{threees}, we defined
$| \xi |_\star = \min ( 1, |\xi| + \ve )$.

Finally,  by (\ref{tre})
both integrals in $d \al$ and $d \tal$ give a factor
$\log^\star \ve$. Then,
by   \eqref{reladder}, \eqref{standardcsi} and \eqref{kn} we have
\begin{eqnarray}
& &
\left| \langle J_\ve , W_{+-,n,{\underline m},{\underline {\widetilde
     {m}}}}^{\rm{ladder}} (\ve^{-1} T) \rangle \right|
\nonumber \\
& \leq & \ve^{-d} ( C
\lambda^2 \ve^{-1})^{n+|
\underline m |+  |\underline {\widetilde m}|} \,  \, (\log^\star \ve)^{4} 
\,
   \ve \,
   \int d\xi  \, dv \,   \left\|
\widehat J_\ve \left( \xi  ,
     \cdot  \right) \right\|_\infty
\left| \widehat f \left(  \f {v} \ve + \f {\xi-2P}  {2 \ve}
   \right) \right| \nnn \\ & & \times \left|
\widehat f \left(\f {v} \ve- \f {\xi-2P} {2 \ve}   \right)  \right|
  \f {\langle v + \xi /2 \rangle \langle v - \xi /2 \rangle} {|
\xi |_\star} \; . \nnn \\ \label{1stappr}
\end{eqnarray}
We focus on the integral factor. Using \eqref{obscond} we get
\begin{eqnarray} & &
\int d\xi  \, dv \,   \left\|
\widehat J_\ve \left( \xi  ,
     \cdot  \right) \right\|_\infty
\left| \widehat f \left( \f {v} \ve+ \f {\xi-2P}  {2 \ve}
   \right) \right| \left|
\widehat f \left( \f {v} \ve - \f {\xi-2P} {2 \ve} \right)  \right|
  \f {\langle v + \xi /2 \rangle \langle v - \xi /2 \rangle} {|
\xi |_\star} \nnn \\ & \leq &\| J\|
  \sup_\xi \left[ \f 1  {| \xi |_\star}
\int dv \, \left| \widehat f \left( \f {v} \ve+ \f {\xi-2P}  {2 \ve}
   \right) \right| \left|
\widehat f \left(\f {v} \ve - \f {\xi-2P} {2 \ve}  \right)  \right|
\langle v + \xi /2 \rangle \langle v - \xi /2 \rangle \right]
\nnn \\
& \leq & \ve^d  \| J\| \| f \|_{H^2 (\erre^d)}^2
  \sup_\xi \left[ \f 1  {| \xi |_\star} \left( \sup_v
\f { \langle v + \xi / 2 \rangle
  \langle v - \xi / 2 \rangle}
   {\left\langle  \f {v} \ve+ \f {\xi-2P}  {2 \ve}
    \right\rangle^2 \left\langle  \f {v} \ve - \f {\xi-2P}  {2 \ve}
    \right\rangle^2 } \right) \right]\; .
\label{prelimest}
\end{eqnarray}

By \eqref{lrangle} one has
\begin{equation}
\f { \langle v + \xi / 2 \rangle
  \langle v - \xi / 2 \rangle}
   {\left\langle  \f {v} \ve + \f {\xi-2P}  {2 \ve}
   \right\rangle \left\langle  \f {v} \ve - \f {\xi-2P}  {2 \ve}
    \right\rangle} \ \leq \ C \langle P \rangle^2
\label{un}
\end{equation}
  and
\begin{eqnarray}
\left\langle  \f {v} \ve + \f {\xi-2P}  {2 \ve}
   \right\rangle \left\langle  \f {v} \ve - \f {\xi-2P}  {2 \ve}
    \right\rangle 
& \geq &  \left\langle \f {\xi-2P}  {2 \ve} \right\rangle
\label{deux}
\end{eqnarray}
Elementary calculations show
\be \label{xinob}
     \sup_\xi \f 1 {\left\langle  \f {\xi-2P}  {2\ve} \right\rangle
  | \xi |_\star} \leq \frac{C}{|P|_\star}.
\ee
Then, plugging \eqref{un}, \eqref{deux} and \eqref{xinob}
  into \eqref{prelimest}
we obtain
\begin{equation*}
\begin{split}
\int d\xi  \, dv \,   \left\|
\widehat J_\ve \left( \xi  ,
     \cdot  \right) \right\|_\infty &
\left| \widehat f \left( \f {v} \ve+ \f {\xi-2P}  {2 \ve}
   \right) \right| \left|
\widehat f \left(\f {v} \ve - \f {\xi-2P} {2 \ve}  \right)  \right|
  \f {\langle v + \xi /2 \rangle \langle v - \xi /2 \rangle} {|
\xi |_\star} \\
\leq & \; C \ve^d \| J\|
\f{\langle P \rangle^2} { | P |_\star}
  \| f \|_{H^2 (\erre^d)}^2.
\end{split} \end{equation*}
By \eqref{1stappr}, recalling that $\lambda = \sqrt \ve$ 
we conclude the proof.
\end{proof}

\begin{remark}
For $n \geq 2$ one can gain another $\ve$ factor in estimate
\eqref{letzte1}. This is easily accomplished
using inequality
$$
\sup_{\theta, \widetilde{\theta}} \int_{\erre^d} \frac {L(k) \, dk}
{\left|  \theta - \Phi_\pm (p,k) + i \eta
\right| | \widetilde \theta - \Phi_\pm (u,k) - i \eta
|}  \leq  \frac {C(\log^\star \eta)^2} { | p - u |_\star}
$$
for estimating the integral in $k_1$ in
\eqref{purenlad}.
\end{remark}

\section{Terms with immediate recollisions only}\label{sec:main}
In this section we prove Theorem 1.1, i.e.
we compute
the contribution of terms consisting of immediate recollisions only,
namely terms with $n=0$ in \eqref{limsup}. The corresponding Feynman
diagrams
are illustrated in Fig. 2.
We stress that, according to  the results in the previous sections,
  these are the sole
non vanishing terms. 

\begin{figure} \label{fig2}
\begin{center}
{}{}\scalebox{0.60}{\includegraphics{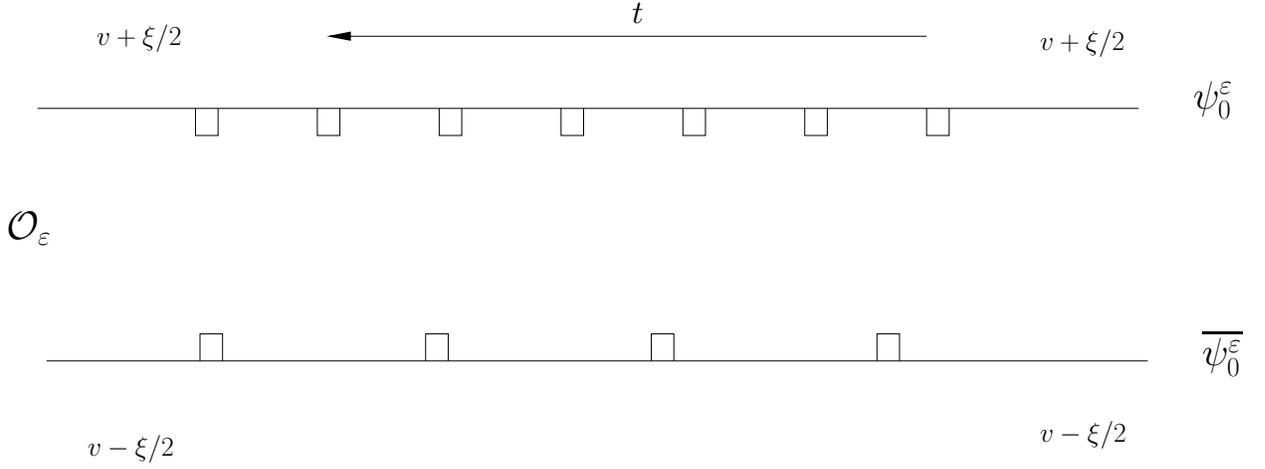}}
\caption{An example of Feynman graph associated to a
pure recollision term. Here $m= 7$, $\widetilde m = 4$.
}
\end{center}
\end{figure}

\begin{proof}
Since the vectors $\underline m$ and $\underline{\widetilde m}$ are
one-dimensional, we can simply denote by $m$ both $\underline m$ and
its only component $m_0$, and by  $\widetilde m$
both $\widetilde {\underline m}$ and $\widetilde{m}_0$.

We preliminarily observe that, according to \eqref{ladderterm},
\eqref{defladderW}, \eqref{initnond},
and the definition of non-diagonal terms given in Section
\ref{sec:non-diag},
for any $t > 0$ and any $\eta>0$ 
the function $ W_{+-,0,{m},{\widetilde m} }^{\rm{ladder}} (t)$ can be
expressed as follows:
\begin{eqnarray}
& &  \widehat W_{+-,0,{m},{\widetilde m} }^{\rm{ladder}} (t; \xi, v)
\nnn \\
                      & = & \lambda^{2 m + 2 \widetilde m}
\left( {2 \pi}   \right)^{-2-d/2} e^{2t\eta} \ve^{-d}
e^{2i \f {v \cdot Q} \ve} \widehat f \left( \f {v} \ve + \f {\xi-2P} {2 
\ve}  \right)
{\ov{\widehat f \left( \f {v} \ve - \f {\xi-2P} {2 \ve}  \right)}}
\nnn \nonumber \\ & & \times \int_\erre d\al \, e^{-it\al} \,
R^{m +1} (\alpha, v+\xi/2, -i\eta) {\Upsilon}^{m}_\eta  (\alpha,
v+\xi/2) \,
\nnn \nonumber \, \int_\erre d\tal \, e^{it\tal} \,
R^{{\widetilde m} +1} (\widetilde \alpha, v-\xi/2, i\eta) \nnn \\ & & \times
{\ov {\Upsilon}}^{\widetilde m}_\eta
  (\widetilde \alpha, v-\xi/2 ) \nnn \\ \label{resprop}
\end{eqnarray}
\begin{eqnarray}
     & = & \lambda^{2 m + 2 \widetilde m}
\left( \sqrt{2 \pi}   \right)^{-d} \ve^{-d}
e^{2i \f {v \cdot Q} \ve} \widehat f \left(\f {v} \ve+ \f {\xi-2P} {2 \ve} 
\right)
{\ov{\widehat f \left( \f {v} \ve - \f {\xi-2P} {2 \ve} \right)}}
\nnn \nonumber \\ & & \times \int_{[0,t]^{2m+1}} \left(\prod_{j=0}^{2m}
   ds_j \right) \delta \left( t -\sum_{j=0}^{2m} s_j \right) \, \left(
   \prod_{j=0}^{m} e^{-i s_{2j} e(v + \xi / 2)} \right) \nnn \\ & & \times
  \left[ \prod_{j=1}^{m} \left( \sum_{\sigma \in \{ \pm \}} \int dk \,
      L(k,\sigma) \, e^{-i s_{2j-1} [e(v + \xi / 2 + k) + \sigma \omega
        (k)]}
\right) \right] \nnn
  \\ & & \times \int_{[0,t]^{2\widetilde {m}+1}} \left(\prod_{j=0}^{2 
\widetilde{m}}
   d\widetilde{s}_j \right) \delta \left( t -\sum_{j=0}^{2 \widetilde{m}}
\widetilde{s}_j  \right)
\,
  \left( \prod_{j=0}^{\widetilde{m}} e^{-i \widetilde{s}_{2j}e(v -
      \xi / 2) } \right) \nnn \\ & & \times
  \left[ \prod_{j=1}^{\widetilde {m}} \left( \sum_{\sigma \in \{ \pm \} } 
\int dk \,
      L(k,\sigma) \, e^{-i \widetilde{s}_{2j-1} [e(v -
      \xi / 2 + k) + \sigma \omega (k)] } \right) \right]
  \nnn
\\ \label{expprop}
\end{eqnarray}
where the functions $R$ and $\Upsilon_\eta$
were defined in \eqref{r-upsilon}. The expression at the r.h.s of
\eqref{resprop} is said the resolvent form, the one in \eqref{expprop}
is said the propagator form of the ladder Wigner function. One can
easily obtain the first from the second one using the identity
$$ \delta \left( t - \sum_{j=0}^{2m} s_j \right) \ = \ \f{e^{\eta t}} {2
    \pi} \int_\erre e^{-i \alpha t} e^{i (\alpha + i \eta)
    \sum_{j=0}^{2m} s_j} \, d\alpha, $$
integrating in the variables $s_j$, and proceeding analogously for the
variables $\widetilde s_j$.

Moreover, we introduce the following quantity:
\begin{equation} \label{fi} \begin{split}
\Phi_P \ : = & \lim_{\eta \rightarrow 0+}
\sum_{\sigma = \pm} \int \frac { L (k,\sigma) \, dk }
{ e (P) - e (P + k) - \sigma \omega (k)  + i \eta} = \Upsilon_{0+} (e(P),P).
\end{split}\end{equation} 
The existence of the limit has been proven in Lemma 4.1 of \cite{e}.

Besides, from the definition of the Boltzmann collision kernel
(see (1.27) in \cite{e}), the total cross section  $\sigma_P$
for an electron with momentum $P$ in the phonon field reads
\be \label{sigmap}
\sigma_P  \ = \ 2\pi \sum_{\sigma \in \{ \pm \}} \int dU \,
  L(P-U) \delta (e(P) - e(U) - \sigma \omega (P-U) )
\ee
and identity (4.12) in \cite{e} yields
\be \label{imphi}
{\mbox {Im}}  \; \Phi_P \ = \ - \f {\sigma_P} 2
\ee

We organize the proof into four steps.


\bigskip

{\em Step 1.} Here we prove
\begin{equation}
\left|  \langle
J_\ve,   W_{+-,0,{m},,{\widetilde m} }^{\rm{ladder}} (t)
\rangle \right| \leq \frac { ( C \lambda^2 t )^{m + \widetilde m}} { m !
\, \widetilde m !} \;. \label{step1}
\end{equation}
Let us mention that
this estimate is stronger than (10.4)
in \cite{e}, but applies only to the ladder terms.

We use the exponential representation of the
propagator (\ref{expprop}). Integrating the delta function with
respect to the variable $s_{2m}$ one gets
\begin{eqnarray*}
& &
\int_{[0,t]^{2m+1}} \left(\prod_{j=0}^{2m}
   ds_j \right) \delta \left( t -\sum_{j=0}^{2m} s_j \right) \, \left(
   \prod_{j=0}^{m} e^{-i s_{2j} e(v + \xi / 2)} \right) \nnn
\\ & & \times
  \left[ \prod_{j=1}^{m} \left( \sum_{\sigma \in \{ \pm \}} \int dk \,
      L(k,\sigma) \, e^{-i s_{2j-1} [e(v + k + \xi / 2) + \sigma \omega
      (k)]} \right) \right] \nnn
\\ & = &  e^{-i t e(v + \xi / 2)} \int_0^t ds_0 \int_0^{t-s_0} ds_2 \dots 
\int_0^{t -
   \sum_{j=0}^{m-2} s_{2j}} ds_{2m-2} \nnn \\ & & \times
\int_0^{t -
   \sum_{j=0}^{m-1} s_{2j}} ds_1
\left( \sum_{\sigma \in \{ \pm \}} \int dk \,
      L(k,\sigma) \, e^{-i s_{1} [e(v + \xi / 2 + k) + \sigma \omega
        (k)
- e(v + \xi / 2 )]} \right)
\dots \nnn \\ & & \dots \times
  \int_0^{t - \sum_{j=0}^{m-1} s_{2j}-\sum_{j=0}^{m-2} s_{2j+1}} ds_{2m - 
1}
\left( \sum_{\sigma \in \{ \pm \}} \int dk \,
      L(k,\sigma) \, e^{-i s_{2m-1} [e(v + \xi / 2 + k) + \sigma \omega
        (k) - e(v + \xi / 2)]} \right)
\end{eqnarray*}

\n
Now we can use inequality \eqref{e-4-7}
and obtain
\begin{eqnarray*}
& &
\left| \int_{[0,t]^{2m+1}} \left(\prod_{j=0}^{2m}
   ds_j \right) \delta \left( t -\sum_{j=0}^{2m} s_j \right) \, \left(
   \prod_{j=0}^{m} e^{-i s_{2j} e(v + \xi / 2)} \right) \right.
\nnn \\ &  & \times \left.
  \left[ \prod_{j=1}^{m} \left( \sum_{\sigma \in \{ \pm \}} \int dk \,
      L(k,\sigma) \, e^{-i s_{2j-1} [e(v + \xi/2 + k) + \sigma \omega
        (k) ]} \right) \right]
\right| \nnn
\\ & \leq  &   \int_0^t ds_0 \int_0^{t-s_0} ds_2 \dots \int_0^{t -
   \sum_{j=0}^{m-2} s_{2j}} ds_{2m-2} \left( \prod_{j=1}^m \int_0^{+
     \infty} ds_{2j - 1} \frac C {\langle  s_{2j - 1} \rangle^{d/2}}
   \right) \nnn
\end{eqnarray*}

\n
Since the space dimension is at least three,
the quantity at the r.h.s. can be estimated by
\begin{eqnarray*}
C^m   \int_0^t ds_0 \int_0^{t-s_0} ds_2 \dots \int_0^{t -
   \sum_{j=0}^{m-2} s_{2j}} ds_{2m-2} & = & \f {C^m t^m} { m !}
\end{eqnarray*}
An analogous estimate holds for the factor in (\ref{expprop}) involving
  the variables $\widetilde s_0, \dots, \widetilde s_{\widetilde m}$.
\begin{eqnarray*}
\left| \langle
  J_\ve,   W_{+-,0,{m},,{\widetilde m} }^{\rm{ladder}} (t)
\rangle \right| & \leq & \frac { (C \lambda^2 t )^{m + \widetilde m}}
{ m ! {\widetilde m} !} \ve^{-d}
\int d \xi \, dv \, | \widehat J_\ve (\xi, v) |
\left| \widehat f \left(\f    v  \ve + \f {\xi - 2P} {2 \ve}
  \right) \right| \left| \widehat f \left(\f v \ve - \f {\xi - 2P} {2
   \ve}  \right) \right|
\nnn \\
  & \leq & \frac { (C \lambda^2 t )^{m + \widetilde m}}
{ m ! {\widetilde m} !} \int d \xi \, \sup_v \left|
  \widehat J_\ve (\xi, v) \right| \int dv \,
\left| \widehat f \left(v +\f {\xi - 2P} {2 \ve}   \right) \right|
  \left| \widehat f \left(v-\f {\xi - 2P} {2
   \ve}  \right) \right|
\end{eqnarray*}
where we performed the change of variables $v \longrightarrow \ve
v$. Estimating the integral in $v$ by Cauchy-Schwarz inequality
and using
\eqref{obscond} we prove \eqref{step1}.

\bigskip
{\em Step 2.}
Here we prove
\begin{equation}
  \lim_{\ve \rightarrow 0} \left| \langle J_\ve,
  W_{+-,0,{m},,{\widetilde m} }^{\rm{ladder}} (\ve^{-1}T)
\rangle \ - \
  \langle  J_\ve,
  {\mathcal W}_{+-,0,{m},{\widetilde m} }^{\rm{ladder}} (\ve^{-1}T)
\rangle \right| ~ = ~ 0
\label{step3}
\end{equation}
where
\begin{eqnarray}
  {\mathcal W}_{+-,0,{m},{\widetilde m} }^{\rm{ladder}}
(t; \xi , v) & :
= &
\left( {2 \pi}\right)^{-2-d/2} \ve^{-d} \lambda^{2m + 2 \widetilde m}
  \Phi^m_P \ov {\Phi^{\widetilde m}_P}
e^{2i \f {v \cdot Q} \ve}
\widehat f \left( \f {v} \ve + \f {\xi-2P} {2 \ve}  \right)
{\ov {\widehat f \left( \f {v} \ve - \f {\xi-2P} {2 \ve}  \right)}}
\nnn \nonumber \\ & & \times e^{2 t \eta} \int_\erre d\al \, e^{-it\al} \,
R^{m +1} (\alpha,v+ \xi/2, -i\eta)  \,  \int_\erre d\tal \, e^{it\tal} \,
{R}^{{\widetilde m} +1}  (\widetilde \alpha, v-\xi/2, i\eta)
\nnn \\
\label{scriptW}
\end{eqnarray}
and $ \Phi_P$ was defined in \eqref{fi}.
Comparing formula \eqref{scriptW} with \eqref{resprop}, one
easily notices that ${\mathcal W}_{+-,0,{m},{\widetilde m} }^{\rm{ladder}}$ equals
${W}_{+-,0,{m},{\widetilde m} }^{\rm{ladder}}$ apart from
the replacement of $\Upsilon_\eta$ by $\Phi_P$.
Then, in the case $m = \widetilde m = 0$ 
there is nothing to
prove. We give the proof in detail for the generic case with both $m$
and $\widetilde m$ greater than zero, and leave to the reader the
case in which either $m$ or $\widetilde m$ vanishes.
We have
\begin{eqnarray}
& &  \langle J_\ve,
{W}_{+-,0,{m},{\widetilde m} }^{\rm{ladder}} (t)
\rangle -  \langle J_\ve,
{\mathcal W}_{+-,0,{m},{\widetilde m} }^{\rm{ladder}} (t)
\rangle \nnn \\
& = & \left( {2 \pi}\right)^{-2-d/2} \ve^{-d} \lambda^{2m + 2 \widetilde 
m}
e^{2 \eta t} \int d\xi \, dv \,  \ov{\widehat J_\ve \left( \xi
   , v \right)} \,
e^{2i \f {v \cdot Q} \ve} \widehat f \left(\f {v} \ve+
  \f {\xi-2P} {2 \ve} \right)
{\ov {\widehat f \left(  \f {v} \ve - \f {\xi-2P} {2 \ve} \right)}}
\nnn \nonumber
\\ & & \times
\int d\al \,  d\tal \, e^{-it\al} \, e^{it\tal} \,
R^{m +1}  \, {\widetilde R}^{{\widetilde m} +1} \left[ {\Upsilon}_\eta^m
{\widetilde {\Upsilon}}_\eta^{\widetilde m} -   \Phi^m_P \ov 
{\Phi_P^{\widetilde m}}
\right] \nnn \\ \label{difference}
\end{eqnarray}
where for brevity we used the notations
\begin{equation*} \begin{split}
R  =  R (\alpha, v + \xi/2, - i \eta), & \qquad
{\Upsilon}_\eta =  {\Upsilon}_\eta (\alpha,
  v + \xi/2)\\
\widetilde R  =  R (\widetilde \alpha, v - \xi/2,  i \eta), & \qquad
\widetilde {\Upsilon}_\eta = \ov{\Upsilon}_\eta (\widetilde \alpha,
  v - \xi/2) \; .
\end{split} \end{equation*}
By the first estimate in \eqref{useful} we have
\begin{eqnarray}
\left|   {\Upsilon}_\eta^m
{\widetilde {\Upsilon}}_\eta^{\widetilde m} -   \Phi^m_P \ov
{\Phi_P^{\widetilde m}}
\right| & = &   \left| {\Upsilon}_\eta^m    \left(  {\widetilde {\Upsilon
      }}_\eta - \ov \Phi_P \right)
\sum_{j=0}^{\widetilde m-1}  {\widetilde {\Upsilon
       }}_\eta^j \, \ov \Phi_P^{\widetilde{m}-j-1} +  \ov 
{\Phi}_P^{\widetilde
m}
\left(  { {\Upsilon
       }}_\eta - \Phi_P \right) \sum_{j=0}^{m-1}  {\Upsilon
       }_\eta^j  \, \Phi_P^{m-j-1} \right|
\nnn \\ & \leq & C^{m + \widetilde m} \left( \left|   {\Upsilon
       }_\eta - \Phi_P \right| +  \left|   {\widetilde {\Upsilon
       }}_\eta - \ov \Phi_P \right| \right) \; .
\label{3}
\end{eqnarray}
Now we estimate (\ref{3}). From the third formula in \eqref{useful}
one has
\begin{eqnarray}
  \left|   {\Upsilon}_\eta - \Phi_P \right|& \leq & C \eta^{-1/2} \left(
| \alpha - e (P) | + | v + \xi / 2  - P | + \eta \right) \nnn \\ &
\leq &
C \eta^{-1/2} \left(
| \alpha - e (v + \xi /2) | + | e (v + \xi / 2) - e (P) |
+ | v + \xi / 2  - P
| + \eta \right) \label{4a}
\end{eqnarray}
Moreover, the first estimate in \eqref{useful} shows that
\begin{equation}
  \left|   {\Upsilon}_\eta - \Phi_P \right| \ \leq \  \left|
    {\Upsilon}_\eta \right| + \left| \Phi_P \right| \ \leq \ C \label{4b}
\end{equation}
From (\ref{4a}) and (\ref{4b}) we obtain
\begin{eqnarray}
  \left|   {\Upsilon}_\eta - \Phi_P \right|
& \leq &  C \eta^{-1/2} \Big( | e (v + \xi / 2) - e (P)
      | + | v + \xi / 2  - P | + \min \left( | \alpha - e
(v + \xi /2) |, 1 \right) \Big)
\label{5}
\end{eqnarray}
where we used $\eta = \ve < 1$.
In order to obtain an analogous estimate 
for $\left|  \widetilde{ {\Upsilon}}_\eta - \ov 
\Phi_P \right| $ we first observe that, due to the symmetry of the
functions $e(\cdot)$, $\omega (\cdot)$, $L(\cdot, \sigma)$, one has
$$ \ov{\Phi}_P = \ov{\Upsilon}_{0+} (e(P),P) = \ov{\Upsilon}_{0+}
(e(P),-P).$$ 
Hence, mimicking \eqref{4a}, \eqref{4b}, \eqref{5}, one obtains
\begin{eqnarray}
  \left| \widetilde  {\Upsilon}_\eta - \ov{\Phi}_P \right|
& \leq &  C \eta^{-1/2} \Big( | e (v - \xi / 2) - e (P)
      | + | v - \xi / 2  + P | + \min \left( | \alpha - e
(v - \xi /2) |, 1 \right) \Big)
\label{5b}
\end{eqnarray}

From \eqref{difference}, \eqref{3}, \eqref{5}, \eqref{5b} we obtain
\begin{eqnarray}
& & \left|  \langle J_\ve,  {W}_{+-,0,{m},{\widetilde m} }^{\rm{ladder}} 
(\ve^{-1}T)
\rangle -  \langle J_\ve,
{\mathcal W}_{+-,0,{m},{\widetilde m} }^{\rm{ladder}} (\ve^{-1}T)
\rangle \right| \nnn \\
& \leq & \ve^{-d} \eta^{-1/2}
(C\lambda^2)^{m + \widetilde m}
\int d\xi \, dv  \left| \widehat J_\ve \left( \xi , v
   \right)  \right|
\left| \widehat f \left( \f {v} \ve+ \f {\xi-2P} {2 \ve}  \right) \right|
  \left| \widehat f \left( \f {v}
\ve - \f {\xi-2P}
{2 \ve}  \right) \right|\nnn \nonumber \\ & & \times
\int_{\erre^2} d\al \,  d\tal \,
| R |^{m +1}  \, | \widetilde R |^{{\widetilde m} +1}
  \Big[ | e (v + \xi / 2) - e (P)| + | e (v - \xi / 2) - e (P)
      |  \nnn \\ & &
+ | v + \xi / 2  - P |
+  | v - \xi / 2  - P |
  + \min \left( | \alpha - e
(v + \xi /2)|, 1 \right)  + \min \left( | \widetilde \alpha - e
(v - \xi /2)|, 1 \right)\Big] \nnn \\
& = & (I) + (I') + (II) + (II') + (III) + (III')
\nnn
\\ \label{6}
\end{eqnarray}
where the last decomposition is made according to
the terms in square parentheses in the r.h.s. of (\ref{6}).

The first term is estimated as follows
\begin{eqnarray*}
  (I) \nnn & \leq & \ve^{-d}
  \eta^{-1/2}
(C \lambda^2)^{m +  \widetilde m} \int d\xi \, \sup_v | \widehat J_\ve
(\xi, v) | \int  dv \,
\left| \widehat f \left( \f {v} \ve+ \f {\xi - 2P} {2\ve }  \right) 
\right|
\left| \widehat f \left( \f {v} \ve- \f {\xi- 2P} {2 \ve }  \right) 
\right|
\nnn \nonumber \\ & & \times
| e ( v + \xi / 2 ) - e (P) |
\int_{\erre^2} d\al \,  d\tal \,
| R |^{m +1}  \, |\widetilde R |^{{\widetilde m} +1} \; .
\end{eqnarray*}

\n
Recalling that $m \ge 1$, we have
\begin{equation*}
\int \f {d \al} {| \al - e (v + \xi / 2 ) + i \eta|^{m+1}}
\ \leq
\ C   \eta^{-m}
\end{equation*}
and analogously for the integral in $\tal$.

By Cauchy-Schwarz
  the integral in $v$ gives
\[ \begin{split} &
\int dv \,  \left| \widehat f \left( \f {v} \ve+ \f {\xi - 2P} {2\ve }
\right) \right|
\left| \widehat f \left( \f {v} \ve- \f {\xi- 2P} {2 \ve }  \right) 
\right|
| e ( v + \xi / 2 ) - e (P) |
\\ \leq & \, \ve^{d/ 2} \| \widehat f\| \left[\int dv \,  \left| \widehat 
f
\left( \f {v} \ve+ \f {\xi - 2P} {2\ve }
  \right) \right|^{2} | e ( v + \xi / 2) - e (P) |^2
\right]^{\f 1 2}
\end{split} \]

Changing variable to $u : = \f v \ve + \f \xi {2 \ve} - \f P \ve$  one
gains a further factor $\ve^{d/2}$. Besides,
using hypothesis \eqref{decdisprel} one gets
$$  | e ( v + \xi / 2) - e (P) | \ = \ | e ( \ve u + P) - e (P) | \
\leq \ C \ve \langle u \rangle^2 \langle P \rangle $$
Then
\[ \int dv \,  \left| \widehat f \left(  \f {v} \ve + \f {\xi - 2P} {2\ve 
}
\right) \right|
\left| \widehat f \left( \f {v} \ve - \f {\xi- 2P} {2 \ve }  \right) 
\right|
| e ( v + \xi / 2 ) - e (P) |
\ \leq \ C \ve^{d+1} \langle P \rangle \| f \|_{H^2 (\erre^d)}
\]
After getting rid of the integral in the variable $\xi$ by \eqref{obscond}
we obtain
$$  
(I) \ \leq \ \ve   \eta^{-1/2} (C \lambda^2 \eta^{-1})^{m + \widetilde m}
  \langle P \rangle
  \| f \|_{H^2 (\erre^d)}  \ \leq \ C^{m + \widetilde m}
  \ve^{1/2} \langle P \rangle \| f \|_{H^2 (\erre^d)}
$$ 
where we chose $\eta = \ve$ and used $\lambda = \sqrt \ve$.
 
Term
$(II)$ in \eqref{6} has the same structure as $(I)$, apart from the
replacement of $e( v +
\xi /2  ) - e (P)$ with $v + \xi / 2 - P$.

For estimating $(III)$
  we proceed as in lemma 10.3 in \cite{e}. Using the
notation $\theta (s) = \min ( |s|,1)$, we observe that
\begin{equation*}
\frac {\theta (\alpha - e (v + \xi / 2))}{|\alpha - e (v + \xi / 2) +
   i \eta|} \ \leq \ \f {C} {\langle \alpha - e (v + \xi / 2)\rangle}
\end{equation*}

Then,
by assumption \eqref{obscond}, hypothesis $m \geq 1$,
Cauchy-Schwarz inequality and the
following two estimates, easily derived from \eqref{uno},
\begin{eqnarray*}
\int \f {d\al } {\langle \al - e (v + \xi / 2) \rangle
| \al - e (v + \xi / 2) + i \eta |^m} & \leq & C^m \eta^{1-m}
\\ \int
\f {d\tal} {| \tal - e (v - \xi / 2) + i \eta |^
{\widetilde {m} + 1}}  & \leq & C^{\widetilde m} \eta^{-\widetilde m}
\end{eqnarray*}
we conclude
\begin{eqnarray*}
(III) & \leq &  \ve^{-d} (C \lambda^2)^{m + \widetilde m} \eta^{-1/2}
\int d\xi \, dv \,  \left| \widehat J ( \xi , v )
\right| \left| \widehat f \left(\f v \ve+ \f {\xi -
       2P} {2 \ve}  \right) \right|
\left| \widehat f \left( \f v \ve- \f {\xi -
       2P} {2 \ve}  \right) \right| \nnn \\ & & \times \int
\f {d\al \, d\tal} {\langle \al - e (v + \xi / 2) \rangle
| \al - e (v + \xi / 2) + i \eta |^m | \tal - e (v - \xi / 2) + i \eta |^
{\widetilde m + 1}} \nnn \\ & \leq & \ve^{\f 1 2} C^{m + \widetilde m}
  \end{eqnarray*}

For terms $(I'), (II'), (III')$ in \eqref{6}
we have the same estimates found for the corresponding terms without
prime.
  Then,
\begin{eqnarray*}
\left| \langle J_\ve,
  W_{+-,0,{m},,{\widetilde {m}} }^{\rm{ladder}} (\ve^{-1}T)
\rangle \ - \
  \langle  J_\ve,
  {\mathcal W}_{+-,0,{m},{\widetilde m} }^{\rm{ladder}} (\ve^{-1}T)
\rangle \right| & \leq &
  C^{m + \widetilde {m}} \sqrt \ve  \langle P \rangle \| f \|_{H^{2} 
(\erre^d)}
\end{eqnarray*}
and (\ref{step3}) is proven.

\medskip

\n
{\em Step 3.}
Here we prove
\begin{equation} \label{step4}
  \langle J_\ve,
{\mathcal W}_{+-,0,{m},{\widetilde m} }^{\rm{ladder}} (\ve^{-1}T)
\rangle \ = \ \f { ( - i T \Phi_P )^m } { m !}
  \f { (i T {\ov{\Phi_P}} )^{\widetilde m} } {{\widetilde m} !}
  \langle J_\ve, W_{+-, {\rm{free}}} (\ve^{-1}T) \rangle
\end{equation}
First, notice that the integrals in $d\alpha$ and $d \widetilde
\alpha$
in the definiton \eqref{scriptW} of
${\mathcal W}_{+-,0,{m},{\widetilde m} }^{\rm{ladder}} (t)$
can be
explicitly computed using residue theorem. One obtains
\begin{equation*}
\int d\al \, \f { e^{-it\al}} {[\alpha - e (v + \xi / 2 ) + i \eta
]^{m+1}} \ = \ - 2 \pi i e^{-\eta t} e^{-it e(v + \xi / 2 )} \f
{(-it)^m}{m!}
\end{equation*}
and analogously  for the integral in $d\widetilde \alpha$.
Therefore, choosing $\eta = \ve$, $t = \ve^{-1} T$, $\lambda = \sqrt \ve$,
from \eqref{scriptW} one obtains
\begin{eqnarray}
  \langle J_\ve,
{\mathcal W}_{+-,0,{m},{\widetilde m} }^{\rm{ladder}} (\ve^{-1}T)  \rangle
& = & \ve^{-d}
  \left( {2 \pi}\right)^{-d/2}  \f
{(-i T \Phi_P)^m}{m!}   \f
{(i T {\ov {\Phi_P}})^{\widetilde m}}{{\widetilde m}!}
\int d\xi \, dv \ e^{-i\f T \ve [e(v + \xi / 2 )-
e(v - \xi / 2 )] } \,\nnn \\
& & \times
  \ov{\widehat J_\ve \left(\xi, v \right)} \,
e^{2i \f {v \cdot Q} \ve } \widehat f \left( \f v \ve + \f {\xi - 2P}
   {2\ve }   \right)
{\ov {\widehat f \left(  \f v \ve - \f {\xi - 2P}
   {2\ve }\right) }}
\nnn \nonumber . \\
  \label{limit4}
\end{eqnarray}

To compute $W_{+-, {\rm{free}}} (\ve^{-1}T)$ we write
$$\gamma_{{\rm e},\ve^{-1}T ,+-, {\rm{free}}} \ = \ e^{-i\f T \ve H_{\rm{e}}} |
     \psi_{0,+}^\ve\rangle \langle  \psi_{0,-}^\ve | e^{i\f T \ve H_{\rm{e}}}$$
or equivalently, in the Fourier space
$$\widehat{\gamma}_{{\rm e}, \ve^{-1}T,+-, {\rm{free}}} (p,u) \ = \ \ve^{-d} 
e^{-i\f T \ve [e
     (p) - e (u) ]} e^{i (p + u ) \cdot \f Q \ve} \widehat f \left(
     \f{p -P} \ve \right) \ov{ \widehat f \left(
     \f{u + P} \ve \right)}. $$
Finally,
  \eqref{step4} immediately follows from \eqref{limit4} by using
  \eqref{freeWt}.

\medskip

\n
{\em Step 4.}
Here we prove
\eqref{result}.
By formulas \eqref{lim}, \eqref{limsup}, \eqref{ladderterm},
\eqref{defladderW} for non-diagonal terms, and eliminating 
terms with $n > 0$ by Lemma \ref{ladd+-},
one gets
\begin{eqnarray*} & &\lim_{\ve \rightarrow 0} \left|
\lim_{L \to \infty} \langle J_\ve, W_{+-} (\ve^{-1} T)
\rangle ~ - ~ e^{- {T} \sigma_P }
\langle J_\ve, W_{+-,\rm{free}} (\ve^{-1} T)
\rangle\right| \nnn
\\ & = &
\lim_{K \to \infty}
\lim_{\ve \rightarrow 0} \left| \sum_{m, \widetilde {m} = 0}^{\left[ \f
{K -1} 2 \right]}
\langle J_\ve, W_{+-,0,{m},{\widetilde {m}} }^{\rm{ladder}}
  (\ve^{-1} T) \rangle ~ - ~ e^{- {T} \sigma_P }
\langle J_\ve, W_{+-,\rm{free}} (\ve^{-1} T)
\rangle\right|,
\end{eqnarray*}
where we used \eqref{step1} to exchange the limits in $K$ and in $\ve$.
Then, from \eqref{step3} and \eqref{step4}
\begin{eqnarray*} & &\lim_{\ve \rightarrow 0} \left| \lim_{L \to \infty}
\langle J_\ve, W_{+-} (\ve^{-1} T)
\rangle ~ - ~ e^{- {T} \sigma_P }
\langle J_\ve, W_{+-,\rm{free}} (\ve^{-1} T)
\rangle\right| \nnn \\
& = &
\lim_{K \to \infty}
\lim_{\ve \rightarrow 0} \left| \sum_{m, \widetilde {m} = 0}^{\left[ \f
{K -1} 2 \right]} \f {(-i T \Phi_P)^m} {m!}  \f {(i T \ov{\Phi_P})^
{\widetilde {m}}} {\widetilde {m} !}
\langle J_\ve,  W_{+-,\rm{free}} (\ve^{-1} T)
\rangle ~ - ~ e^{- {T} \sigma_P }
\langle J_\ve, W_{+-,\rm{free}} (\ve^{-1} T)
\rangle\right| \nnn \\ & \leq &
  \limsup_{K \to \infty}
  \left| e^{- {T} \sigma_P }  ~ - ~ \sum_{m, \widetilde {m} = 0}^{\left[ \f
{K -1} 2 \right]} \f {(-i T \Phi_P)^m} {m!}
   \f {(i T \ov{\Phi_P})^
{\widetilde {m}}} {\widetilde {m} !} \right|  \limsup_{\ve \rightarrow 0}
  \left|\langle J_\ve, W_{+-,\rm{free}} (\ve^{-1} T)
\rangle\right| 
\end{eqnarray*}
The first factor in the r.h.s. vanishes due to \eqref{imphi},
while the second factor is bounded since
$$
    \left|\langle J_\ve, W_{+-,\rm{free}} (t)
\rangle\right| = \Big|\; {\rm Tr} \; \mc{O}_\ve^\star 
|e^{-itH_{\rm{e}}}\psi_{0,+}^\ve\rangle\langle
  e^{-itH_{\rm{e}}}\psi_{0,-}^\ve| \; \Big| \leq \| \mc{O}_\ve\|.
$$
  The proof is complete.
\end{proof}


\section{Appendix}

Here we prove Lemma \ref{threees}.
\begin{proof}
Inequality \eqref{uno} corresponds
to
formula (5.19) in \cite{e}, and we refer to that paper for the proof.
  We prove \eqref{due}. We treat the case of
$\Phi_+$, since for $\Phi_-$ the proof is the same.

Consider a tiling of the $k$-space in cubes $\{ Q_i \}_{i \in \natu}$
of size  thinner than
$\widetilde \rho$ defined in \eqref{intersection}, \eqref{transversality}.
  Fix $l_0 : =
[ \log^\star (\widetilde \rho \eta^{-1})]$,
and define the sets
\be \begin{split} \nnn
S_l \ : = & \{ k \in \erre^d : e^{-l-1} \widetilde \rho \leq
| \theta - \Phi_+ (p,k) | \leq  e^{-l} \widetilde \rho \}, \ \ 1 \leq l 
\leq
l_0 - 1 \\
S_{l_0} \ : = &  \{ k \in \erre^d :
| \theta - \Phi_+ (p,k) | \leq  e^{-{l_0}} \widetilde \rho \} \\
S_0 \ : = &  \{ k \in \erre^d :
| \theta - \Phi_+ (p,k) | \geq  \widetilde \rho / e \}
\end{split}\ee
and analogously the sets $\widetilde S_{\widetilde l}$,
$0 \leq \widetilde l \leq l_0$, where $\widetilde\theta$
replaces $\theta$ in the definitions above.

\n
Then,
\begin{eqnarray}
  \int_{\erre^3} \frac {\mc{L}(k) \, dk}
{\left|  \theta - \Phi_+ (p,k) + i \eta
\right| | \widetilde \theta - \Phi_+ (u,k) - i \eta
| \langle p \rangle \langle u \rangle}
& \leq &
\sum_i \sum_{l,{\widetilde l} = 0}^{l_0}
  U_{i, l, \widetilde l}
\left| Q_i \cap
S_l \cap {\widetilde S}_{\widetilde l} \right|
\nnn \\ \label{splitted}
\end{eqnarray}
where by $| \cdot |$
we denoted the Lebesgue measure in $\erre^d$
and introduced the notation
$$
   U_{i,l,\widetilde l} := \sup_{k \in Q_i \cap
S_l \cap {\widetilde S}_{\widetilde l}} \left( \f {  {\mathcal L} (k) }
{\left|  \theta - \Phi_+ (p,k) + i \eta
\right| | \widetilde \theta - \Phi_+ (u,k) - i \eta|
\langle p \rangle \langle u \rangle} \right)
$$
Since the actual value of $\widetilde \rho$ does not play any role, in
what follows we will absorb it in the constant $C$.

We prove the following estimate:
\begin{equation} \begin{split}
   U_{i, l, \widetilde l}
  \ \leq \
C \f{e^{l+ \widetilde l}} {\langle \theta \rangle^{\f 1 2}
   \langle \widetilde \theta \rangle^{\f 1 2} }
\sup_k \left[ \langle k \rangle^4 |\mc{ L} (k)| \right] , \qquad 0 \leq
l, \widetilde l \leq l_0.
\label{everyl}
\end{split} \end{equation}
First, we treat the case $1 \leq l, \widetilde{l} \leq l_0$. We have
\begin{equation*} \begin{split}
U_{i,l,\widetilde l}  \ \leq &  \sup_{k \in Q_i \cap
S_l \cap {\widetilde S}_{\widetilde l}} \left( \f { 1}
{\left|  \theta - \Phi_+ (p,k) + i \eta
\right| | \widetilde \theta - \Phi_+ (u,k) - i \eta|} \right) \,
  \sup_{k \in Q_i \cap
S_l \cap {\widetilde S}_{\widetilde l}} \left( \f { |\mc{ L}  (k) |}
{\langle p \rangle \langle u \rangle} \right)
\end{split} \end{equation*}
For the factor $\left|  \theta - \Phi_+ (p,k) + i \eta
\right|^{-1}$ we proceed as follows
\begin{itemize}
\item if $1 \leq l < l_0$ then we apply
$ \left|  \theta - \Phi_+ (p,k) + i \eta
\right|^{-1} \leq C e^{l}   $;
\item if $l = l_0$ then we use
\begin{equation*}
{\left|  \theta - \Phi_+ (p,k) + i \eta
\right|^{-1} \ \leq \ | \eta |^{-1} \ \leq \ C
e^{l_0 } }.
\end{equation*}
\end{itemize}
Analogously,
\be \label{neq0}
| \widetilde \theta -
\Phi_+ (u,k) - i \eta|^{-1} \ \leq \ C e^{\widetilde l}, \qquad
1 \leq \widetilde l  \leq \widetilde l_0.
\ee
We finally obtain
\begin{equation}
\label{zero}
   \sup_{k \in Q_i \cap
S_l \cap {\widetilde S}_{\widetilde l}} \left( \f { 1}
{\left|  \theta - \Phi_+ (p,k) + i \eta
\right| | \widetilde \theta - \Phi_+ (u,k) - i \eta|} \right)
\ \leq \ C e^{l + \widetilde l}, \qquad 1 \leq l, \widetilde l \leq l_0.
\end{equation}

Now we estimate $\sup_{k \in Q_i \cap
S_l \cap {\widetilde S}_{\widetilde l}} \left( \f { | L (k) |}
{\langle p \rangle \langle u \rangle}\right)$.
Using inequalities \eqref{lrangle}, \eqref{decdisprel} we have
\begin{eqnarray} \label{p-2}
  \langle p \rangle^{-2} \ \leq \ C \f { \langle k \rangle^{2}} {
\langle e (p+k) \rangle} \ \leq \ C \f { \langle k \rangle^{2}} {
\langle \theta -  \Phi_+ (p,k) - \theta + \om (k)
\rangle}   & \leq & C \f {
\langle k \rangle^{4}
\langle  \theta -  \Phi_+ (p,k) \rangle}
{  \langle  \theta  \rangle}.
\end{eqnarray}
Notice that in $S_l$, $1 \leq l \leq l_0$  one has
$\langle  \theta -  \Phi_+ (p,k)  \rangle \leq 1 + e^{-l}
\widetilde \rho$, therefore such a quantity can be estimated
by a constant  and
we end up with
\begin{equation} \label{p-1}
  \langle p \rangle^{-1} \ \leq \ C  \f {
\langle k \rangle^{2}} {  \langle  \theta  \rangle^{1/2}}, \, \qquad
\mbox{$k\in S_l$, \  $1 \leq l
\leq l_0$}.
\end{equation}
Analogously,
\begin{equation} \label{u-1}
  \langle u \rangle^{-1} \ \leq \ C  \f {
\langle k \rangle^{2}} {  \langle  \widetilde
\theta  \rangle^{1/2}}, \, \qquad
\mbox{$k\in {\widetilde S}_{\widetilde l}$,  $1 \leq {\widetilde l}
\leq l_0$}.
\end{equation}
Therefore, from \eqref{zero}, \eqref{p-1}, \eqref{u-1} one obtains
\eqref{everyl} for $l, \widetilde l \neq 0$.

To estimate $U_{i,0,0}$ we observe that, proceeding like in
\eqref{p-2} and using
\begin{equation*}
\frac{\langle  \theta -  \Phi_+ (p,k) \rangle}{| \theta -
\Phi_+ (p,k) + i \eta |} \ \leq \ C, \qquad k \in S_0,
\end{equation*}
one gets
\begin{equation}
  \frac 1
{\left|  \theta - \Phi_+ (p,k) + i \eta
\right|  \langle p
\rangle}
  \ \leq \ C  \f {
\langle k \rangle^{2}} {  \langle  \theta  \rangle^{1/2}},\, \qquad
k\in S_0. \label{egy}
\end{equation}

Analogously,
\begin{equation} \label{ketto}
  \frac 1
{\left|  \widetilde \theta - \Phi_+ (u,k) - i \eta
\right|  \langle u
\rangle}
  \ \leq \ C  \f {
\langle k \rangle^{2}} {  \langle  \widetilde \theta  \rangle^{1/2}},
\, \qquad
k\in \widetilde S_{0} .
\end{equation}
Then, from \eqref{egy} and \eqref{ketto} one obtains \eqref{everyl} for
$l, \widetilde l = 0$.
Finally, to obtain \eqref{everyl} for $l=0$, $ \widetilde l \neq 0$ it is
sufficient to use inequalities
\eqref{u-1}, \eqref{egy} and \eqref{neq0}, and analogously for the
case $l \neq 0$, $\widetilde l = 0$.

The volume factor
$|Q_i \cap S_l \cap
{\widetilde S}_{\widetilde l}|$ in \eqref{splitted} is estimated as 
follows:
\begin{itemize}
\item if both $l$ and $\widetilde l$ are greater than zero, then we
   use
inequality \eqref{transversality}
   and estimate the volume of
   integration by
$$ \left| Q_i \cap
S_l \cap {\widetilde S}_{\widetilde l} \right|
\ \leq \
C \f{e^{-l- \widetilde l}} {| p - u |};$$
\item if $l=0$ and $\widetilde l \neq 0$,
then we use inequality \eqref{intersection} and obtain
$$  \left| Q_i \cap
S_0 \cap {\widetilde S}_{\widetilde l} \right|
\ \leq  \ C e^{- \widetilde l} ;$$
vice versa, if $l > 0, \widetilde l=0$, then $ \left| Q_i \cap
S_l \cap {\widetilde S}_{0} \right|$ is estimated
by $C e^{- l} $.
\item If both $l$ and $\widetilde l$ equal zero, then the volume
  $ \left| Q_i \cap
S_0 \cap {\widetilde S}_{0} \right|$
   is estimated by $C$.
\end{itemize}

Then,
\begin{eqnarray*} & &
\sum_i \sum_{l,{\widetilde l} = 0}^{l_0}  \int_{Q_i \cap
S_l \cap {\widetilde S}_{\widetilde l}} \frac { \mc{ L}(k) \, dk}
{\left|  \theta - \Phi_+ (p,k) + i \eta
\right| | \widetilde \theta - \Phi_+ (u,k) - i \eta|
\langle p \rangle \langle u \rangle}
\nnn \\
& \leq & C \, 
 \f{ 1 + 2 l_0 + l_0^2/|p-u| } {\langle \theta
   \rangle^{\f 1 2} \langle \widetilde \theta
   \rangle^{\f 1 2}}
\sum_i{\|\langle \cdot \rangle^4  \mc{ L}
  \|_{L^\infty (Q_i)}}
\ \leq \  C\f {
   (\log^\star \eta)^2} {{\langle \theta
   \rangle^{\f 1 2}  \langle \widetilde \theta
   \rangle^{\f 1 2} } \min (1,
|p-u|)}   \sum_i \|\langle \cdot \rangle^4  \mc{ L}  \|_{L^\infty (Q_i)}
\nnn 
\end{eqnarray*}
The decay property \eqref{qdecay} of $ \mc{ L}$ guarantee that
the sum in $i$ is finite. Then,
\begin{equation} \label{pu}
  \int_{\erre^d} \frac { \mc{ L}(k) \, dk}
{\left|  \theta - \Phi_+ (p,k) + i \eta
\right| | \widetilde \theta - \Phi_+ (u,k) - i \eta|
\langle p \rangle \langle u \rangle} \ \leq \
   C\f {
   (\log^\star \eta)^2} {{\langle \theta
   \rangle^{\f 1 2}  \langle \widetilde \theta
   \rangle^{\f 1 2} } \min (1,
|p-u|)}
\end{equation}

\n
Inequality \eqref{pu}
can be improved in the case $| p - u | < \eta$. Indeed, in this case,
we can estimate terms with $l, \widetilde l > 0$ by means of
inequality (\ref{intersection})
in spite of \eqref{transversality}. We obtain
\begin{equation} \begin{split} &
  \int_{\erre^d} \frac { \mc{ L}(k) \, dk}
{\left|  \theta - \Phi_+ (p,k) + i \eta
\right| | \widetilde \theta - \Phi_+ (u,k) - i \eta| \langle p \rangle
  \langle u \rangle} \\
   \leq  &  C \, 
 \f{ 1 + 2 l_0 + l_0 e^{l_0}} {\langle \theta
   \rangle^{\f 1 2} \langle \widetilde \theta
   \rangle^{\f 1 2}}
\sum_i{\|\langle \cdot \rangle^4  \mc{ L}
  \|_{L^\infty (Q_i)}} \  \leq \ C
  \f {\log^\star \eta} { \langle \theta \rangle^{\f 1 2} \langle
  \widetilde \theta \rangle^{\f 1 2} \eta}
\label{sumll}
\end{split} \end{equation}

\n
Combining \eqref{pu} and \eqref{sumll}  we arrive at \eqref{due}.

Estimate \eqref{tre} is trivial.

\end{proof}

\end{document}